\documentclass[manuscript, nonacm]{acmart}
\usepackage{bbm}
\usepackage{mathtools}
\usepackage{algorithm}
\usepackage{algpseudocode}
\usepackage{graphicx}
\usepackage{textcomp}
\usepackage{xcolor}
\usepackage{cleveref}
\usepackage{multirow}
\usepackage{multicol}
\usepackage{subcaption}
\usepackage{comment}
\usepackage{tikz}
\usepackage{pifont}
\usepackage{threeparttable}
\usepackage{makecell}

\usepackage{braket}
\usepackage{float}
\usepackage{listings}
\usepackage{xcolor}
\usepackage[newfloat]{minted}

\DeclareMathOperator*{\argmax}{arg\,max}

\AtBeginDocument{%
  }

\setcopyright{acmlicensed}
\copyrightyear{2018}
\acmYear{2018}
\acmDOI{XXXXXXX.XXXXXXX}

\acmJournal{JACM}
\acmVolume{37}
\acmNumber{4}
\acmArticle{111}
\acmMonth{8}

\newcommand{\revA}[1]{#1}
\newcommand{\revB}[1]{#1}
\newcommand{\revC}[1]{#1}

\begin{document}

\title[SkipOPU]{SkipOPU: An FPGA-based Overlay Processor for Large Language Models with Dynamically Allocated Computation}

\author{Zicheng He}
\email{hzc1997ece@gmail.com}
\orcid{0000-0003-1598-2118}
\authornotemark[1]
\affiliation{%
  \institution{University of California, Los Angeles}
  \city{LA}
  \state{CA}
  \country{USA}
}

\author{Anhao Zhao}
\authornote{Equal contribution.}
\affiliation{%
 \institution{ Institute of Digital Twin, Eastern Institute of Technology}
 \city{Ningbo}
 \country{China}
}

\author{Xiaoyu Shen}
\authornotemark[2]
\affiliation{%
 \institution{Ningbo Key Laboratory of Spatial Intelligence and Digital Derivative, Institute of Digital Twin, Eastern Institute of Technology}
 \city{Ningbo}
 \country{China}
}
\email{xyshen@eitech.edu.cn}

\author{Chen Wu}
\authornotemark[2]
\affiliation{%
  \institution{Chiplet CAD and Manufacturing Engineering Research Center of Zhejiang Province, Ningbo Institute of Digital Twin, Eastern Institute of Technology}
  \city{Ningbo}
  \country{China}}
\email{cwu@idt.eitech.edu.cn}

\author{He Lei}
\authornote{Corresponding authors.}
\affiliation{
  \institution{Eastern Institute of Technology}
  \city{Ningbo}
  \country{China}}
\email{lei.hexun@gmail.com}
\renewcommand{\shortauthors}{Zicheng He and Anhao Zhao, et al.}

\begin{abstract}
Large language models (LLMs), \revC{particularly decoder-only models with auto-regressive decoding}, have achieved remarkable performance across a wide range of tasks, but their inference efficiency remains a critical bottleneck due to rapidly growing parameters \revC{and memory-bound generation phases}.
Recent advances in \emph{dynamic computation allocation} address this challenge by exploiting the highly uneven contributions of different tokens and layers, enabling selective execution that significantly reduces redundant computation while preserving model accuracy.
However, existing hardware platforms and accelerators are primarily optimized for uniform, static execution, limiting their ability to efficiently support such dynamic inference patterns.
In this work, we propose \textbf{SkipOPU}, an FPGA-based overlay processor that dynamically allocates computation across tokens and layers with high flexibility through a lightweight routing mechanism. 
First, we decouple reduction operations from element-wise computation in nonlinear modules and perform reductions incrementally. 
This design enables both stages to be fused with adjacent linear operations (router or matrix multiplication), effectively hiding nonlinear latency within the pipeline.
Second, motivated by asymmetric sensitivity to numerical precision between activation and weight, we design a processing element (PE) array that efficiently supports \emph{float–fixed} hybrid execution.
A novel DSP overpacking technique is introduced to maximize hardware utilization while minimizing resource overhead. 
Finally, we develop a proactive on-chip KV \revB{(Key-Value)} history buffer that exploits cross-layer KV invariance of pruned tokens,  eliminating irregular HBM \revB{(High Bandwidth Memory)} accesses during decoding and supplementing off-chip bandwidth through high-locality on-chip reuse.
Experimental results demonstrate that \textbf{SkipOPU} on an AMD U280 FPGA outperforms GPU and other FPGA-based accelerators by $1.31\times - 2.55\times$ in bandwidth efficiency for LLMs inference with dynamic computation allocation.
Moreover, cross-layer KV reuse reduces up to $25.4\%$ KV storage overhead across varying sequence lengths. 

\end{abstract}



\maketitle
\section{INTRODUCTION}
Large language models (LLMs) have demonstrated unprecedented capabilities across a wide range of applications, including natural language processing~\cite{FewShotLearner, BERT, OPT, LLM2Vec}, computer vision~\cite{ViT, MovieGen}, and recommendation systems~\cite{BERT4Rec, TransAct}. Such remarkable performance is largely driven by the continuous scaling of model capacity~\cite{kaplan2020scaling,hoffmann2022training,su2024unraveling}. However, increasing model size also incurs substantial inference-time costs in both computation and memory, a challenge that becomes particularly pronounced in settings involving multimodal inputs and multi-turn, agentic inference~\cite{belcak2025small,fan2025visipruner,song2025smallthinker,liu2026vica}. 
As a result, deploying LLMs places significant pressure on memory bandwidth, storage capacity, and compute resources, especially on resource-constrained platforms. These challenges are further exacerbated by the growing demand for local deployment, motivated by requirements for offline inference, improved energy efficiency~\cite{EfficientLLM}, and stronger data privacy guarantees~\cite{li2024privacylargelanguagemodels, schwinn2023adversarialattacksdefenseslarge}. 

Model pruning has emerged as a widely adopted and effective strategy to reduce computational complexity and memory usage. 
Previous work has extensively explored static pruning, including both unstructured sparsity\cite{ChatOPU, FlashLLM} and structured sparsity\cite{PP-Transformer, NVSM_SPARSE, FlightLLM, 10818746}. 
By leveraging sparse weight patterns, these approaches typically depend on custom accelerator architectures optimized for sparse matrix multiplication (SpMM) to translate theoretical complexity reductions into practical hardware efficiency. 
However, the static nature of these approaches limits adaptability across diverse inputs and often results in noticeable accuracy degradation under aggressive sparsity~\cite{dlp,MoD,SkipGPT}. 

To address this limitation of static pruning, dynamic computation allocation techniques have been proposed to adapt computation at runtime according to input characteristics~\cite{D-LLM}. A prominent line of work focuses on sparse attention mechanisms, which dynamically restrict the attention range during the scaled dot-product (\emph{i.e.} $QK^T$ and $SV$ stages) based on token relevance or structural priors~\cite{ELSA, FACT, SANGER,yuan2025native}. 
More recent studies go beyond attention-scope optimization and explore dynamic allocation of entire computational units, ranging from coarse-grained layers to finer-grained sub-layer components such as self-attention blocks and feed-forward network blocks within a single Transformer layer~\cite{LearnToSkip,wu2025routing,MoD,D-LLM}. In these approaches, each computation unit is equipped with a lightweight router, typically implemented as a small MLP \revB{(Multi Layer Perceptron)}, that decides whether the unit should be executed or skipped for each token. In practice, important tokens are routed through most of the model’s parameters, while less informative tokens can bypass substantial computation. This adaptive execution paradigm closely mirrors human language processing, where semantically critical words receive deeper analysis, while predictable or less informative words are processed more shallowly. 
Among these methods, SkipGPT~\cite{SkipGPT} stands out as a flexible and representative framework. It enables token-wise dynamic execution by independently routing tokens through the Multi Head Attention \revB{(MHA)} and feed-forward network submodules within each layer, thereby providing fully flexible token-level control over computation. This design contrasts with prior approaches that enforce a fixed number of participating tokens per layer block~\cite{MoD} or impose rigid coupling between attention and feed-forward execution~\cite{D-LLM}, allowing SkipGPT to capture richer and more fine-grained token-wise computational dynamics.

Despite achieving state-of-the-art accuracy–efficiency trade-offs under comparable pruning budgets, however, existing evaluations primarily report theoretical reductions in computational complexity. These gains have yet to be systematically validated on real hardware platforms. As a result, how to efficiently map SkipGPT-style dynamic execution onto modern hardware architectures while translating algorithmic savings into tangible end-to-end performance improvements remains a set of significant and largely unexplored challenges~\cite{han2025informed,ding2026llms}.

\begin{figure}
    \centering
    \vspace{0cm}
    \includegraphics[width=0.95\textwidth]{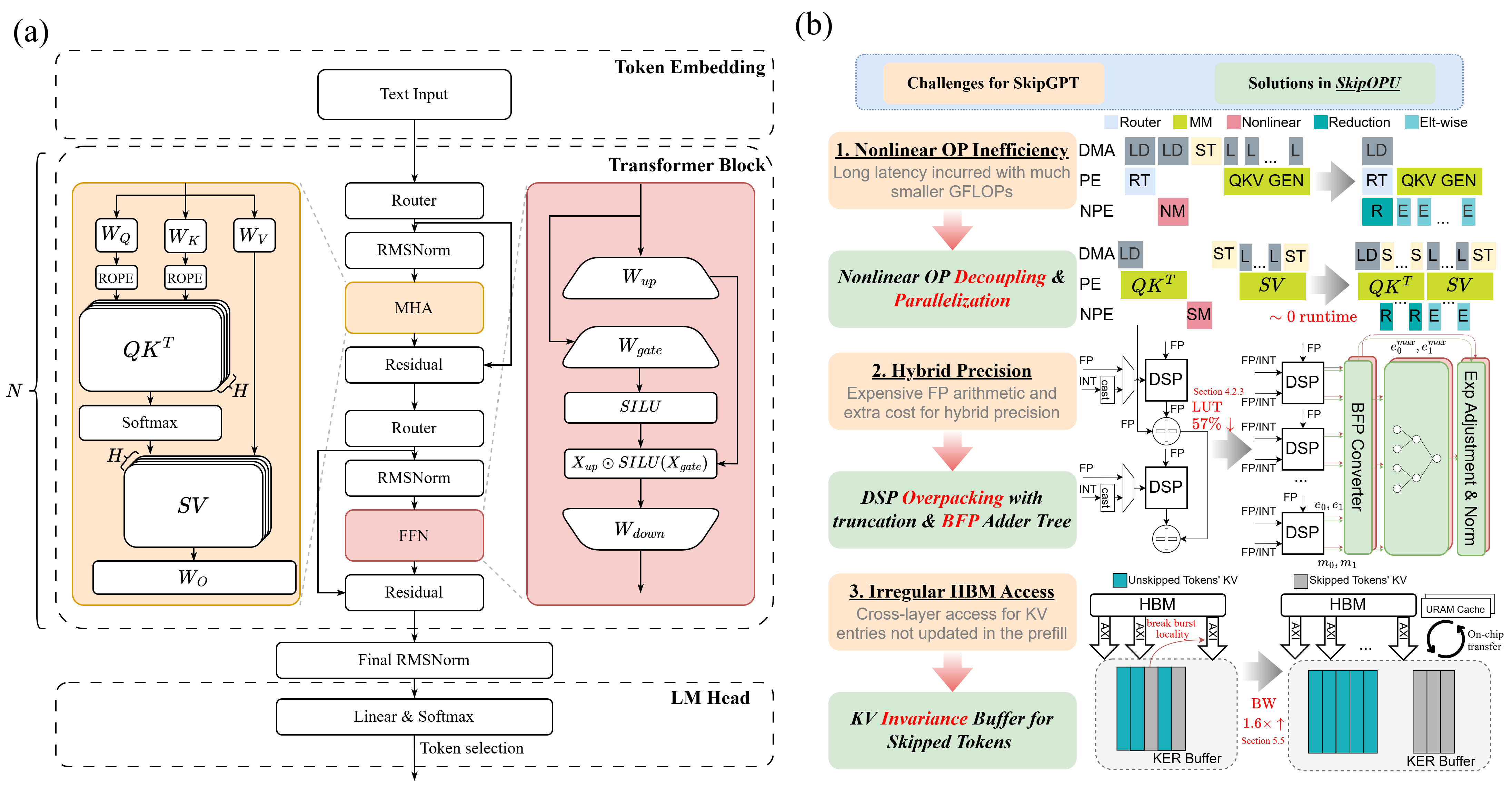}
    \caption{(a) The framework of SkipGPT models. (b) Challenges in SkipGPT computation, and the corresponding solution in \textbf{SkipOPU}.\revB{(DMA: Direct Memory Access, NPE: Nonlinear Processing Engine)} }
    \label{fig:Challenges}\vspace{-1.5em}
\end{figure}

\Cref{fig:Challenges}.(b) summarizes the key challenges in developing hardware accelerators for SkipGPT: 
\textbf{(1) Long latency induced by nonlinear operations and routers. } 
Nonlinear operations typically exhibit limited computational efficiency due to inherent data dependencies, which prevent fully pipelined dataflow and enforce serialized execution.
In SkipGPT, the Gumbel-Softmax-based routing \cite{Gumbel} introduces additional low–arithmetic-intensity operations beyond conventional nonlinearities.
These control-dominant computations are inefficiently supported on conventional GPU architectures, thereby amplifying inference overhead.
As a result, despite contributing a relatively small fraction of total FLOPs, nonlinear operations and routing modules can disproportionately impact end-to-end latency.
\textbf{(2) Hybrid precision preference and under-utilization of DSP-based PEs \revB{(Processing Elements)}.} 
Prior quantization studies, such as \cite{AWQ, GPTQ}, demonstrate that model weights can be aggressively quantized to 4-bit integers with minimal accuracy degradation.
However, activations typically remain in FP16 to preserve numerical fidelity, resulting in a hybrid precision execution regime. 
Existing DSP-based PE arrays neither fully exploit the intrinsic multiply-accumulate capabilities of DSP blocks nor efficiently support mixed-precision computation\cite{LPFP, MP-OPU}.
Consequently, hybrid precision execution leads to under-utilized hardware resources and imposes a substantial computational burden on resource-constrained edge devices.
\textbf{(3) Incompatibility with KV \revB{(Key-Value)} cache optimization strategies.} 
Dynamic layer skipping introduces KV-cache inconsistencies. 
Existing approaches address this issue by optimizing either storage efficiency (\emph{e.g.}, KV eviction) or model fidelity (\emph{e.g.}, partial attention skipping).
A particular compromise is to substitute missing KV entries with those from earlier layers, thereby preserving attention semantics with minimal storage overhead. 
However, such cross-layer KV reuse introduces highly irregular HBM \revB{(High Bandwidth Memory)} access patterns that fragment AXI \revB{(Advanced eXtensible Interface)} burst transactions, severely degrading memory locality and off-chip bandwidth utilization.

To this end, we propose \textbf{SkipOPU}, an FPGA-based overlay processor designed to accelerate LLMs with dynamically allocated computation across both tokens and layers, targeting the SkipGPT framework. 
First, we introduce a novel dataflow architecture that effectively hides the latency of nonlinear operations interleaved between linear layers. 
By decoupling the numerical feature dependencies within Softmax and RMSNorm \revB{(Root Mean Square Layer Normalization)}, \revC{defined as $y_i = \frac{x_i}{RMS(x)}\times \gamma_i$ with $RMS(x) = \sqrt{\epsilon + \frac{1}{n}\sum_{i=1}^nx_i^2}$}, from the traditional sequential datapath, our design enables incremental computation that can be seamlessly fused with streaming outputs from PE arrays, thereby improving pipeline efficiency. 
Second, we develop a mixed-precision processing engine with fine-grained data packing to fully exploit the inherent multiplication and addition capabilities of FPGA DSP blocks. 
Inspired by the Block Floating Point (BFP) methodology \cite{BFP}, we replace costly floating-point accumulations with lightweight fixed-point summation combined with FP–BFP format conversion.
This approach substantially reduces arithmetic overhead while maintaining numerical fidelity.
Finally, we adopt a KV reuse strategy for handling missing entries and observe that the KV values of skipped tokens remain unchanged across layers until reactivated. 
Leveraging this invariance, we partition KV-cache read during the decode by placing pre-known invariant KV entries in on-chip URAM \revB{(UltraRAM)} while retrieving the remaining entries in HBM. 
This separation transforms irregular cross-layer memory accesses into high-locality on-chip transactions, mitigating AXI-level burst fragmentation and effectively supplementing limited off-chip bandwidth.

In summary, our main contributions are:
\begin{itemize}
  \item We present the first hardware system that supports token-wise and layer-wise dynamic pruning, integrated with hybrid-precision execution and KV-aware cache management for LLM inference.
  \item We propose a novel latency-hiding dataflow tailored for nonlinear operations, along with a dedicated nonlinear processing engine (NPE) that supports major LLM nonlinearities(\emph{e.g.} RoPE \revB{(Rotary Positional Embedding)} and SwiGLU \revB{(Swish Gated Linear Unit)}) through tile-wise streaming computation.
  \item We design a DSP-overpacked PE array with BFP-based accumulation to efficiently enable mixed-precision computation while incurring minimal hardware overhead. 
  \item We develop a proactive on-chip KV history buffer that exploits cross-layer KV invariance to convert irregular HBM accesses into high-locality on-chip reuse, significantly improving memory efficiency.
\end{itemize}
Experimental results show that \textbf{SkipOPU} implemented on an AMD U280 FPGA achieves $1.31\times$ - $2.55\times$ higher bandwidth efficiency compared to GPUs and prior FPGA-based accelerators for LLM inference with dynamically allocated computation. 
Furthermore, by applying the proposed KV reuse strategy, \textbf{SkipOPU} reduces KV memory storage requirements by up to 25.4\% across varying input and output sequence lengths.
\section{BACKGROUND AND MOTIVATIONS}
This section reviews the relevant background and motivates our design by identifying key opportunities for improving computation and memory efficiency.
\subsection{SkipGPT framework}
Modern LLMs are typically built with a decoder-only Transformer architecture, formed by stacking multiple identical decoder layers. To start with inference, the entire input prompt is processed in parallel to generate the initial output token, a phase known as the prefill stage. Subsequently, the model enters the multi-step decode stage, where each step consumes the accumulated context to produce the next token in an auto-regressive manner. 

The SkipGPT framework~\cite{SkipGPT} shares the same inference process, but enhances LLMs by introducing a router before each submodule per layer, as shown in \Cref{fig:Challenges}.(a). 
For each token $x$, the router is evaluated prior to the submodule, $f_l$, at $l$-th transformer layer to determine whether the corresponding computation should be executed or skipped. 
In SkipGPT, the router is implemented as a linear projection that produces a two-dimensional categorical distribution, $r_l = \mathbf{W}_\theta^T x \in \mathbb{R}^2$, and the decision is drawn by sampling from the categorical distribution. 
Similar routing mechanisms are employed by other dynamic layer-pruning approaches, including MoD~\cite{MoD}, SkipLayer~\cite{LearnToSkip}, and D-LLM~\cite{D-LLM}, which differ mainly in the specific form of the router but share the same lightweight design principle. 
Specially, the output of the submodule can be written as
\begin{align}
x_{l+1} = \begin{cases}
f_l(x_l)+ x_l, & \text{if sampled result is } 1, \\
x_l, & \text{otherwise}
\end{cases}
\end{align}
While only a subset of tokens is selected to execute MHA at $l$-th layer, it is common practice to preserve attention semantics by allowing each unskipped token to attend to the full token sequence. For such skipped tokens, their key and value vectors can be obtained through different strategies; a bypass approach is to reuse the most recent KV representations from the last layer where their computation was performed~\cite{han2025informed}. As a result, the unskipped attention computation for $n$-th token at $l$-th layer, $x_{l,n}$ can be formulated as 
\begin{align}
\begin{split}
    &\text{Attention}(x_{l,n}) = \text{softmax}\left( \frac{x_{l,n}W_Q^l \cdot K_l^T}{\sqrt{d_k}}\right) \cdot V_l,\text{where } K_l, V_l\in \mathbb{R}^{L\times d_k} \\
\end{split}
\end{align}
where $L$ denotes the sequence length and $d_k$ is the dimensionality of the key and value vectors.
For each position $i$, the key vector is defined as $K_l[i, :] = x_iW_K^l$ if the routing decision for $x_{l,i}$ is equal to 1; 
Otherwise, it is inherited from the most recent executed layer via recursive fallback, i.e., $K_l[i,:] = K_{l-1}[i,:]$. The same rule applies to the value matrix $V_l$.

\subsection{Motivation}
The motivation for \textbf{SkipOPU} arises from the distinctive characteristics of dynamic layer pruning methods, particularly the Gumbel-Softmax-based routing mechanism and the disparity between KV-cache update and reuse.

Although nonlinear operations account for only a small fraction of the total arithmetic operations, they contribute disproportionately to inference latency due to their complex data dependencies and multi-pass computation. 
Although prior designs such as MCoreOPU\cite{MCoreOPU} and \cite{NPE} propose special function units for nonlinear operations to reduce memory traffic for intermediate results, the row-based computation still prevents effective pipelining. 
This challenge is exacerbated by the Gumbel-softmax based routing, which further requires joint dataflow design across linear and nonlinear kernels to enable seamless integration and effective latency hiding. 

Beyond nonlinear operations, the performance of matrix multiplication must also be carefully considered. 
Although architectures such as systolic arrays have been extensively optimized to maximize utilization for dense GEMM \revB{(GEneral Matrix Multiply)}, the intrinsic compute capability of DSP blocks remains underexploited when supporting hybrid-precision workloads
Moreover, conventional floating-point accumulation is costly in both area and power, and its overhead scales with PE-array size, directly limiting the attainable compute density on edge devices. 
Addressing inefficient DSP utilization and expensive floating-point accumulation is therefore critical to unleashing the untapped performance potential. 

Another primary motivation comes from the disparity between KV generation and consumption when applying KV reuse for attention semantics maintenance.
Although only a subset of tokens update their KV entries at each layer, all tokens must still be attended to by unskipped queries, forcing the model to repeatedly reuse KV entries across layers. 
During decoding, this mismatch translates into irregular memory accesses to off-chip HBM. As modern accelerators rely on wide, burst-based AXI transactions to achieve high bandwidth, such random accesses severely degrade spatial locality and collapse effective memory throughput.
Although irregular KV accesses can be mitigated by redundantly storing identical KV entries at every layer, doing so eliminates the memory savings provided by dynamic pruning.
This observation motivates a memory system that preserves the storage benefits of dynamic pruning while restoring burst-friendly, high-locality KV access for efficient attention computation.

\section{DATAFLOW OPTIMIZATION}
\label{sec:dataflow}
This section introduces the dataflow optimization for both the router and the self-attention under KV reuse mechanism to facilitate the latency hiding of nonlinearities lied between.  
\subsection{Fused dataflow for routing and RMSNorm}
To reduce memory traffic among the router, the executed submodule, and the intervening RMSNorm while effectively hiding the latency of nonlinear operations, we introduce a fused dataflow, as illustrated in \Cref{alg:fused_router}.
In a conventional dataflow, routing and normalization require multiple full passes over the activation matrix, resulting in two large off-chip loads and an additional store before the subsequent submodule can begin. 
In contrast, our dataflow leverages the token-wise independence of routing decisions and partitions the activation matrix into on-chip tiles, allowing each slice to be reused by both the RMSNorm and the following submodule without repeated off-chip accesses. 

To further accelerate the multi-pass RMSNorm, we decouple its reduction phase from the element-wise normalization and execute the reduction in parallel with the linear router computation. 
\revB{
Specifically, these concurrent operations are isolated across the PE array (indicated in yellow in \Cref{alg:fused_router}) and the NPE module (indicated in green). Through this dedicated hardware-algorithm co-design, the unified dataflow achieves fully pipelined execution by eliminating structural hazards and pipeline bubbles.
}
As a result, the normalization statistics become available immediately after the routing logits are produced. 
After determining the routing decisions with straight-through argmax, the unskipped tokens together with their precomputed mean and variance are selected directly from the on-chip buffer to complete normalization.
The resulting tile-wise normalization is seamlessly overlapped with the pipelined matrix multiplication of the following submodule, effectively hiding the latency of both the reduction and normalization stages.
For energy efficiency, the normalized activations are written back to the on-chip buffer while being streamed to the PE array for matrix multiplication, overwriting the original unnormalized activations in place to avoid repeated computation.

\begin{algorithm}
\caption{Deep-Fused Router and RMSNorm dataflow}
\label{alg:fused_router}
\begin{algorithmic}[1]
\small
\Require Router parameter $\mathbf{W}_\theta \in \mathbb{R}^{D\times2}$; RMSNorm parameter, $\gamma \in \mathbb{R}^D$; QKV generation weights, $\mathbf{W}_Q, \mathbf{W}_K, \mathbf{W}_V \in \mathbb{R}^{D\times D}$; RoPE weights, $\{\mathbf{R}^d_{\Theta,m} \in \mathbb{R}^{d \times d} | m \in \{1, 2, ..., L\}\}$; Activation matrix, $X \in \mathbb{R}^{L\times D}$ and the block size along the row $B_r$ and column $B_c$ respectively. Tile size along the reduction dimension $S$
\State Divide activation matrix into $T_r$ blocks, $\Set{\mathbf{X}_i \in \mathbb{R}^{B_r \times D} | i = 1, 2, ... T_r}$ and QKV generation weights into $T_c$ blocks. $\Set{\mathbf{W}_Q^i, \mathbf{W}_K^i, \mathbf{W}_V^i \in \mathbb{R}^{D \times B_c } | i = 1, 2, ... T_c}$
\For {$1 \leq i \leq T_r$}
\For {$1 \leq k \leq T_n, T_n = D // S$}
\State Load $X_i^k = X_i[:, kS:(k+1)S] \in \mathbbm{R}^{B_r\times S}$ from DDR to on-chip buffer
\State \hspace*{-\fboxsep}\colorbox{yellow!30}{\parbox{\linewidth-\fboxsep - 3em}{$\text{logit} \leftarrow \text{logit} + X_i^k\mathbf{W}_\theta[kS:(k+1)S, :] \in \mathbb{R}^{B_r\times 2}$}}
\State \hspace*{-\fboxsep}\colorbox{green!30}{\parbox{\linewidth-\fboxsep - 3em}{$\mu \leftarrow \text{RowSum}(X_i^k)/D + \mu$; $var \leftarrow var + \text{RowMean}[(X_i^k)^2] - [\text{RowMean}(X_i^k)]^2 \in \mathbb{R}^{B_r}$}}
\EndFor
\State {$d = \mathbbm{1}(\argmax(\ln(\text{logit}) + \pi), \pi\in G(0, 1)$}
\For {$1 \leq j \leq T_c$}
\For{$1 \leq k \leq T_n$}
\State Load ${X_i^k}'= X_i^k[d==1] \in \mathbb{R}^{B_r'\times S}$ from on-chip buffer; Load $\mathbf{W}_V^{jk}=\mathbf{W}_V^j[kS:(k+1)S, :] \in \mathbb{R}^{S\times B_c}$ from HBM
\If{j == 1}
\State \hspace*{-\fboxsep}\colorbox{green!30}{\parbox{\linewidth-\fboxsep - 6em}{$\mu' = \mu[d==1]$, $\sigma' = \sqrt var[d==1]$; ${X_i^k}' = \frac{{X_i^k}' - \mu'}{{\sigma'}} \times \gamma[kS:(k+1)S]$}}
\State Store ${X_i^k}'$ to on-chip buffer to replace the original ${X_i^k}$
\EndIf
\State \hspace*{-\fboxsep}\colorbox{yellow!30}{\parbox{\linewidth-\fboxsep - 4.5em}{$V_{i,j}^k \leftarrow {X_i^k}'\mathbf{W}_V^{jk} + V_{i,j}^{k-1} \in \mathbb{R}^{B_r'\times B_c}$ \Comment{\textcolor{red}{Only V generation is shown for simplicity}}}}
\EndFor
\EndFor
\EndFor
\end{algorithmic}
\end{algorithm}

\subsection{Fused dataflow for self-attention mechanism}
Following the same decoupling and incremental-computation paradigm, we further fuse the softmax operation with adjacent linear computation, as shown in \Cref{alg:fused_attention}.
\revB{
Maintaining the notation from \Cref{alg:fused_router}, operations mapped to the PE array are indicated in yellow, while miscellaneous non-linear operations executed within NPE module are highlighted in green.
}
The softmax reduction phase—specifically, the computation of row-wise numerical features (i.e., the maximum and exponential summation)—is reformulated using the update rule from FlashAttention\cite{FlashAttention}, allowing these features to be accumulated incrementally within a single pass. 
This incremental formulation allows the concurrent feature updates to be tightly pipelined with the $QK^{T}$ computation, effectively hiding the softmax reduction latency.
Once a full row of intermediate results is available, subsequent matrix multiplication can proceed immediately by loading the value matrices from HBM and the intermediate attention scores from DDR. 
The intermediate tiles are then streamed through the normalization unit and seamlessly consumed by the $SV$ computation, incurring negligible additional overhead.

One subtle yet critical challenge in the fused attention dataflow stems from the head-wise execution of self-attention and the irregular HBM access patterns induced by cross-layer KV reuse for skipped tokens.
The head-wise computation inherently limits arithmetic intensity, pushing the self-attention mechanism toward a memory-bound regime. 
This effect is further exacerbated by fragmented KV accesses across layers, making self-attention predominantly memory-bound in practice even during the prefill stage.
To mitigate this bottleneck, we pack the computation of multiple attention heads within the PE array, thereby aggregating KV accesses to improve spatial locality and recover effective HBM bandwidth. 
However, such a packing produces $QK^{T}$ results from multiple heads concurrently, which would otherwise require substantial on-chip storage to materialize full attention rows under a conventional softmax dataflow.
Leveraging the decoupled softmax formulation, we avoid this overhead by immediately streaming raw $QK^{T}$ tiles to off-chip memory while retaining only the associated numerical features on-chip, enabling multi-head packing without incurring excessive on-chip memory consumption.

\begin{algorithm}[htbp]
\caption{Deep-Fused self-attention dataflow}
\label{alg:fused_attention}
\begin{algorithmic}[1]
\small
\Require Matrices $\left\{ \mathbf{Q}_i | i = 1, 2, ..., H \right\} \in \mathbb{R}^{L'\times d}$, $\left\{ \mathbf{K}_i, \mathbf{V}_i | i = 1, 2, ..., H \right\} \in \mathbb{R}^{L\times d}$, and the block size along the row, $B_r$ and $B_c$ 
\State Divide $\mathbf{Q}_i$ into $T_r$ blocks, $\Set{\mathbf{Q}_i \in \mathbb{R}^{B_r \times d} | i = 1, 2, ... T_r}$ and divide $\mathbf{K}_i, \mathbf{V}_i$ into $T_c$ blocks, $\Set{\mathbf{K}_i, \mathbf{V}_i \in \mathbb{R}^{B_c \times d} | i = 1, 2, ... T_c}$
\For {$1 \leq h \leq H/2$}
\For{$1 \leq i \leq T_r$}
\State Load $\mathbf{Q}_{2h}^i, \mathbf{Q}_{2h+1}^i$ from DDR to on-chip buffer
\For {$1 \leq j \leq T_c$}
\State Load $\mathbf{K}_{2h}^i, \mathbf{K}_{2h+1}^i$ from HBM to on-chip buffer
\State \hspace*{-\fboxsep}\colorbox{yellow!30}{\parbox{\linewidth-\fboxsep - 4.5em}{$\mathbf{S}_{2h}^{ij} = \mathbf{Q}_{2h}^i{\mathbf{K}_{2h}^j}^T$ and $\mathbf{S}_{2h+1}^{ij} = \mathbf{Q}_{2h+1}^i{\mathbf{K}_{2h+1}^j}^T \in \mathbb{R}^{B_r\times B_c}$}}
\State \hspace*{-\fboxsep}\colorbox{green!30}{\parbox{\linewidth-\fboxsep - 4.5em}{$\forall u \in \left\{0,1\right\}$ $\tilde{m}_{2h+u}^{ij} = \text{rowmax}(\mathbf{S}_{2h+u}^{ij}) \in \mathbb{R}^{B_r}$, $\tilde{\mathbf{P}}_{2h+u}^{ij} = \text{exp}(\mathbf\mathbf{S}_{2h+u}^{ij} - \tilde{m}_{2h+u}^{ij})$, $\tilde{\ell}_{2h+u}^{ij} = \text{rowsum}(\tilde{\mathbf{P}}_{2h+u}^{ij}) \in \mathbb{R}^{B_r}$}}

\State \hspace*{-\fboxsep}\colorbox{green!30}{\parbox{\linewidth-\fboxsep - 4.5em}{$\forall u \in \left\{0,1\right\}$ $m_{2h+u}^{i,new} = \text{max}(m_{2h+u}^{i}, \tilde{m}_{2h+u}^{ij}), \mathcal{\ell}_{2h+u}^{i,new} = \text{exp}(m_{2h+u}^{i}-m_{2h+u}^{i,new})\mathcal{\ell}_{2h+u}^{i} + \text{exp}(\tilde{m}_{2h+u}^{ij} - m_{2h+u}^{i,new})\mathcal{\tilde{\ell}}_{2h+u}^{ij}$}}
\State Set $m_{2h+u}^{i} \leftarrow m_{2h+u}^{i,new}, \mathcal{\ell}_{2h+u}^{i} \leftarrow \mathcal{\ell}_{2h+u}^{i,new}$; Store $\mathbf{S}_{2h}^{ij}$ and $\mathbf{S}_{2h+1}^{ij}$ back to off-chip memory
\EndFor
\For {$1 \leq j \leq T_c$}
\State Load $\mathbf{S}_{2h}^{ij}$ and $\mathbf{S}_{2h+1}^{ij}$ from DDR to on-chip buffer; Load $\mathbf{V}_{2h}^{j}, \mathbf{V}_{2h+1}^{j} \in \mathbb{R}^{B_c\times d}$ from HBM to on-chip buffer
\State \hspace*{-\fboxsep}\colorbox{green!30}{\parbox{\linewidth-\fboxsep - 4.5em}{$\forall u \in \left\{0,1\right\}$ $\mathbf{S}_{2h+u}^{ij} \leftarrow exp(\mathbf{S}_{2h+u}^{ij} - m_{2h+u}^i) / \mathcal{\ell}_{2h+u}^i \in \mathbb{R}^{B_r\times B_c}$}}
\State \hspace*{-\fboxsep}\colorbox{yellow!30}{\parbox{\linewidth-\fboxsep - 4.5em}{$\mathbf{O}_{2h}^{i} \leftarrow \mathbf{S}_{2h}^{ij} \mathbf{V}_{2h}^j + \mathbf{O}_{2h}^{i}$, $\mathbf{O}_{2h+1}^{i} \leftarrow \mathbf{S}_{2h+1}^{ij} \mathbf{V}_{2h+1}^j + \mathbf{O}_{2h+1}^{i} \in \mathbb{R}^{B_r\times d}$}}
\EndFor
\State Store  $\mathbf{O}_{2h}^i, \mathbf{O}_{2h+1}^i$ back to DDR
\EndFor
\EndFor
\end{algorithmic}
\end{algorithm}
\section{HARDWARE ARCHITECTURE}
\label{sec:hardware}
\subsection{Overview}
\label{sec:overview}
\begin{figure}
    \centering
    \vspace{0cm}
    \includegraphics[width=0.9\textwidth]{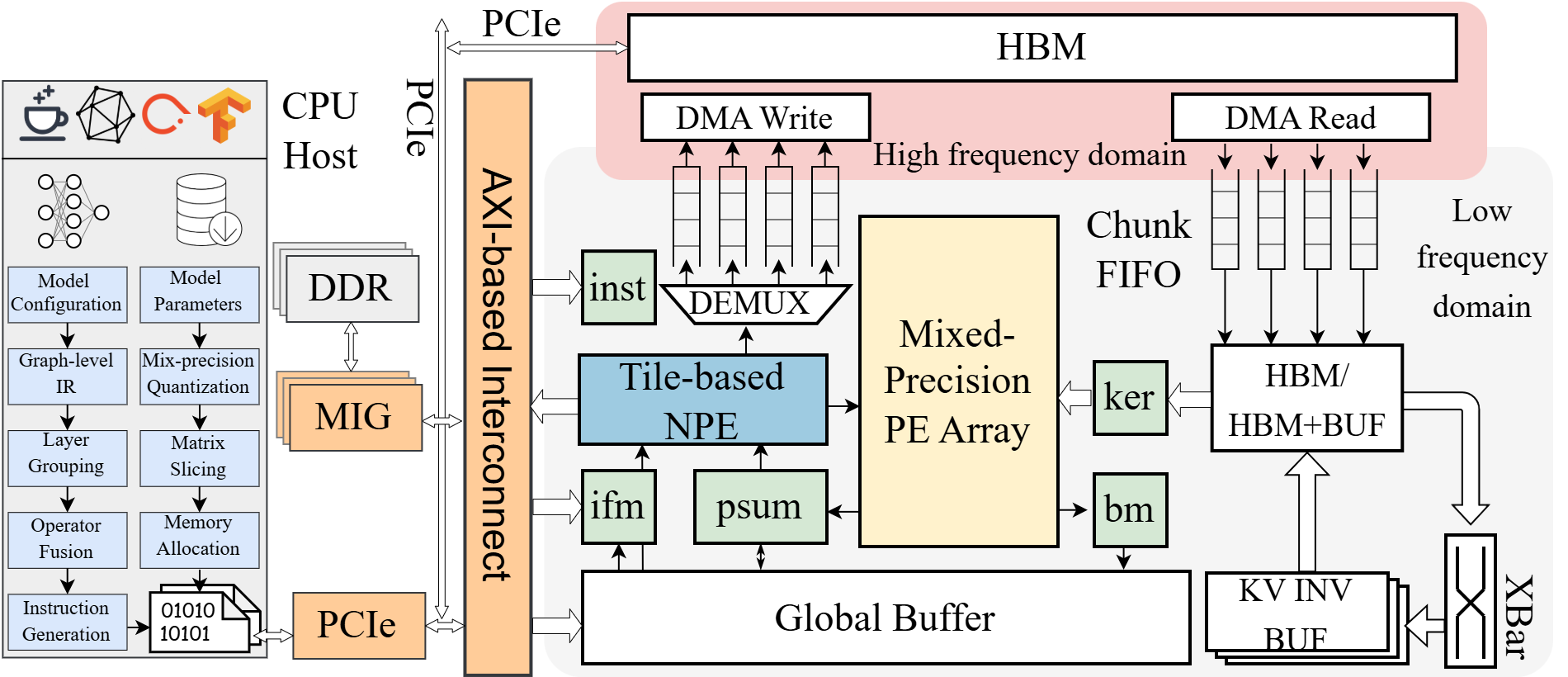}
    \caption{\revB{Architectural overview of the SkipOPU framework with heterogeneous frequency domains. (MIG: Memory Interface Generator)}}
    \label{fig:Overview}\vspace{-1.5em}
\end{figure}

\Cref{fig:Overview} provides the general architecture of the proposed \textbf{SkipOPU} for SkipGPT-style models. 
It contains a CPU host, an FPGA fabric for computation acceleration, and heterogeneous off-chip memories (HBM2 and DDR4) for storing model parameters and intermediate results.
\revC{
On the host side, a dedicated C++ compiler is developed to process SkipGPT-style model configurations and generate the low-level instructions. 
Inheriting its core framework from the established OPU series\cite{OPU}, the compiler transforms the model topology into a graph-level Intermediate Representation (IR), and subsequently performs multiple optimization passes—including operator fusion, weight quantization, and memory allocation—prior to code generation. 
The optimized IR is finally mapped to a series of instructions to control the running process of \textbf{SkipOPU} while the instruction set is further extended to support runtime-dynamic sequence lengths and layer-skipping control flows.
This instruction-driven execution paradigm also enables broad support for dynamic layer-pruning frameworks, such as MoD and D-LLM;
Although these frameworks differ in router configurations and KV cache management strategies, they share the same underlying computational primitives. 
Consequently, their execution behavior can all be mapped onto the overlay architecture on FPGA through software-level compilation updates without requiring structural modifications to the underlying hardware.
}

Communication between the CPU and FPGA is established via PCIe, whereas data transfers between the FPGA and peripheral memories are handled through AXI-based interconnects.
\revC{
To satisfy the intense memory demands of LLMs, the HBM memory interface is configured to operate at its maximum supported frequency on target board, maximizing available off-chip bandwidth. Conversely, the  accelerator core runs at a lower 225MHz frequency due to both physical constraints and workload characteristics. 
Specifically, beyond DSP-based PE array, the accelerator integrates heavy control logic implemented by LUT/FF resources to support the overlay architecture flexibility, making higher operating frequencies increasingly difficult to sustain under FPGA timing-closure and routing constraints.
Moreover, the decoding phase of LLM inference is predominantly memory-bound, where throughput is primarily limited by off-chip bandwidth rather than computation throughput, making a lower core frequency an acceptable and optimized design trade-off. 
To bridge the distinct clock domains between HBM and accelerator core, we deploy Asynchronous Chunk FIFOs realized through Xilinx asynchronous FIFO primitives. 
The Clock Domain Crossing (CDC) mechanism is implicitly managed by this hard-validated IP, which utilizes Gray-coded pointers and multi-stage synchronizers to guarantee safe data synchronization, ensuring a consistent, aligned data stream for the kernel computation.
}

The accelerator is primarily composed of dedicated on-chip buffers and specialized function units to maximize data reuse and hardware utilization.
To sustain a continuous dataflow during PE execution, a set of localized ping-pong buffers---\revC{namely the Input Feature Map (IFM), Kernel (KER), and Partial Sum (PSUM) buffers—are implemented directly using Block RAM (BRAM) to provide the wide, multi-port bus widths required for parallel execution.}
\revC{
A centralized Global Buffer (scratchpad), also built from BRAM, is further introduced to accommodate diverse data reuse patterns across distinct linear projection layers.
Dynamic runtime execution is driven by the Bitmask (BM) module.
}
The Bitmask module consumes router results produced by the PE array to make computation decision and selectively fetches token vectors from the global buffer to drive the execution of subsequent submodules.
Tile-based NPE handles nonlinear operations and can be flexibly inserted into the computation pipeline, either during operand preparation or result offloading. 
\revC{
Intermediate activation matrices are offloaded to off-chip DDR4 memory, which is managed via Xilinx's industry-standard Memory Interface Generator (MIG) IP to guarantee robust and high-efficiency physical layer control.
Computed KV cache entries are routed back to the HBM channels. To maximize memory throughput, a dedicated Demultiplexer (DEMUX) assigns KV cache segments to distinct HBM ports (see details in \Cref{sec:memory_opt}). This mapping enforces localized access alignment, ensuring that each AXI channel reads and writes strictly to its corresponding local HBM channel to prevent cross-channel routing congestion.
}
During decoding, the KV invariance buffer is automatically updated to capture key–value entries that remain unchanged across adjacent layers. 
When self-attention computation is active in the next layer, the buffer outputs are concatenated inline with fetched HBM data, effectively supplementing the off-chip bandwidth.

\subsection{Mixed-precision PE array}
\subsubsection{Overpacking design for FP16 with truncation}
Matrix multiplications on FPGAs are typically mapped to dedicated DSP blocks, whose operand bit-widths are fixed by the underlying hardware. 
For example, on the Xilinx U280 platform, each DSP48E2 unit natively supports a $27\text{-bit} \times 18\text{-bit}$ multiplication. However, LLM inference commonly employs data formats such as INT8 or FP16.
This mismatch between the native width of the DSP operand and the effective precision of LLM workloads leads to systematic underutilization of the available DSP compute capacity, as only a fraction of the arithmetic resources of each DSP are used per operation.

\begin{figure}[htbp]
    \centering
    \vspace{0cm}
    \includegraphics[width=0.9\textwidth]{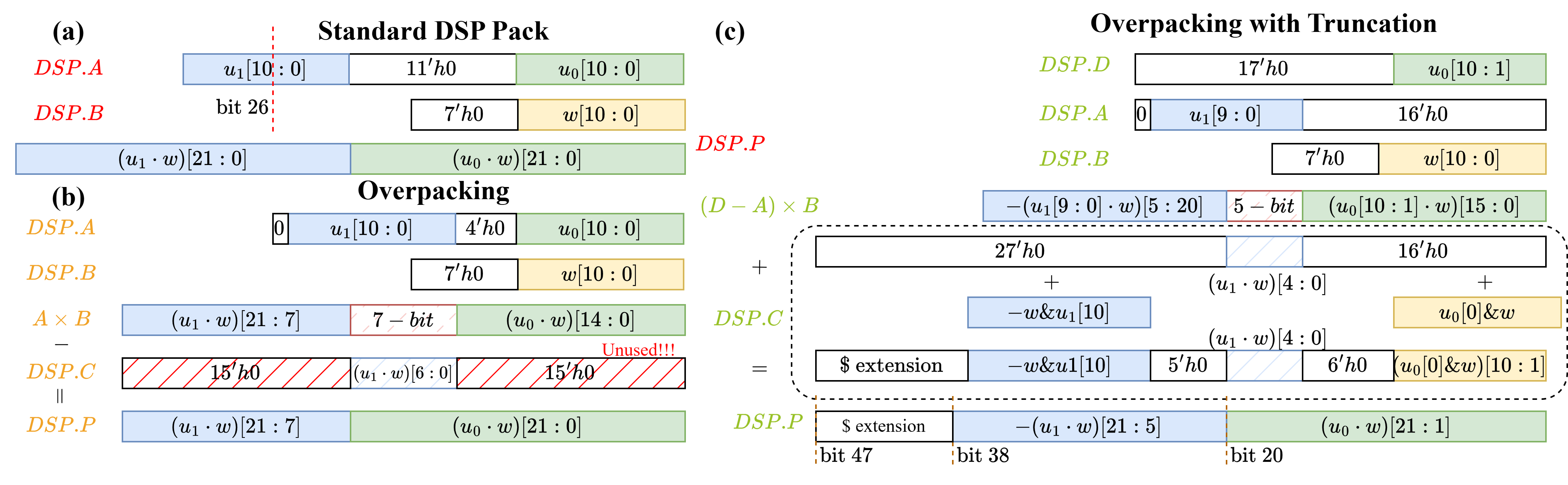}
    \caption{Comparison between differences DSP packing techniques for FP16 fraction multiplication, \revC{where $w, u_0, u_1$ represents the significand of input and weights in FP16 format}.}
    \label{fig:Overpack}\vspace{-1.5em}
\end{figure}

To address this mismatch, DSP packing techniques are commonly employed to better exploit the fixed arithmetic width of FPGA DSP blocks. 
\revA{
Specifically, the input bitwidth is split into multiple pieces, each of which holds low-bitwidth data to improve multiplication parallelism while preserving the correct product \cite{xilinx_whitepaper_2017, xilinx_whitepaper_2020}. 
However, such conventional methods cannot be directly applied to floating-point scenarios.
}
\revC{
As illustrated in \Cref{fig:Overpack}.(a), the fraction calculation of two parallel FP16 multiplication fails to map to the DSP architecture due to the strict operand-width constraints imposed.
Here, $u_0$, $u_1$, and $w$ represent the 11-bit hidden-bit extended significands of the FP16 
operands $W_0, W_1, X$, defined as $u_0 = \{1\text{'b1}, W_0[9:0]\}$, $u_1 = \{1\text{'b1}, 
W_1[9:0]\}$, and $w = \{1\text{'b1}, X[9:0]\}$, respectively.
} 
To relax these constraints, overpacking techniques such as those proposed in \cite{Overpack} intentionally allow intermediate corruption in the DSP multiplication results and subsequently recover the correct outputs through post-processing. 
As shown in \Cref{fig:Overpack}.(b), this method leverages the DSP’s built-in addition path for result recovery, requiring only an additional $\text{7-bit}$ multiplication \cite{FlightVGM}. 

\revA{
Despite these prior packing advancements, they treat multiplication as a monolithic operation, leaving the internal adder resources of the DSP underutilized.
Observing that a multiplication can be structurally decomposed into successive stages of partial additions, we exploit this property to break the rigid precision boundaries. 
By selectively truncating the operand significands, we reduce the effective bit-width prior to execution, which significantly lowers the number of corrupted guard bits and minimizes the subsequent recovery overhead. 
The residual information lost during truncation is then seamlessly concatenated with the corrupted recovery bits and injected back into the datapath via the DSP's native C-port—a technique we define as \textit{information injection}.
}
As illustrated in \Cref{fig:Overpack}.(c), we introduce several key design choices to simplify information injection and corruption recovery within the DSP datapath.
First, we truncate the least significant bit (LSB) of the operand $u_0$ and the most significant bit (MSB) of the operand $u_1$. This truncation pattern is carefully selected to ensure that the omitted bit segments do not introduce unwanted cross-terms during hardware reconstruction. 
\revC{
Formally, by decomposing the packed operands based on their truncated segments, the monolithic packing equation can be expanded as
\begin{equation}
(u_0 - 2^{16}u_1)w = 2(u_0[10:1]\cdot w - 2^{15}u_1[9:0]\cdot w)+ u_0[0] \cdot w +2^{26}(u_1[10]\cdot -w)
\end{equation}
where the term, $u_0[10:1]\cdot w - 2^{15}u_1[9:0]\cdot w$ represents the primary multiplication performed with truncated operand widths.  
The residual terms, $u_0[0] \cdot w$ and $2^{26}(u_1[10] \cdot -w)$, constitute the exact information targeted for out-of-band injection. Because these injected terms are separated by sufficient binary shifting (26 bits), their respective product segments occupy completely disjoint bit-fields. 
}
Second, it is important to note that the corrupted terms introduced by overpacking must be eliminated via subtraction, whereas the information lost due to truncation must be restored via addition, leading to separate arithmetic operations and additional logic.
To avoid this overhead, we invert the sign of the operand $u_1$ through the DSP pre-adder, allowing both corruption cancellation and truncation recovery to be realized through a single additive operation via the DSP C-port.
With this design, only a 5-bit overpacking of the operand $u_1[9:0]$ is required, along with a corresponding 5-bit auxiliary multiplication for corruption recovery, enabling efficient utilization of the DSP’s arithmetic resources with minimal overhead.
\revC{
After post-processing via the inner adder embedded in the DSP primitive, the $21$ MSBs of the fractional product of the first operand pair ($1.X^{\text{mantissa}} \times 1.W_0^{\text{mantissa}}$) can be extracted from the lower-order segment, $DSP.P[20:0]$. Concurrently, the $18$ MSBs of the fractional product corresponding to the second operand pair ($1.X^{\text{mantissa}} \times 1.W_1^{\text{mantissa}}$) are mapped within the higher-order segment, $DSP.P[38:21]$. 
} 

\subsubsection{Mixed-precision PE architecture}
After applying overpacking techniques, each DSP unit can generate multiple partial products per cycle within the same silicon footprint, substantially increasing multiplication throughput. 
This increased arithmetic density, however, places a correspondingly higher demand on accumulation, as more partial results must be combined within each PE. 
Although addition is generally less expensive than multiplication, accumulation in FP16 still incurs nontrivial overhead due to exponent alignment, normalization, and rounding, whose cost scales with the size of the PE array.

\begin{figure}
    \centering
    \vspace{0cm}
    \includegraphics[width=0.95\textwidth]{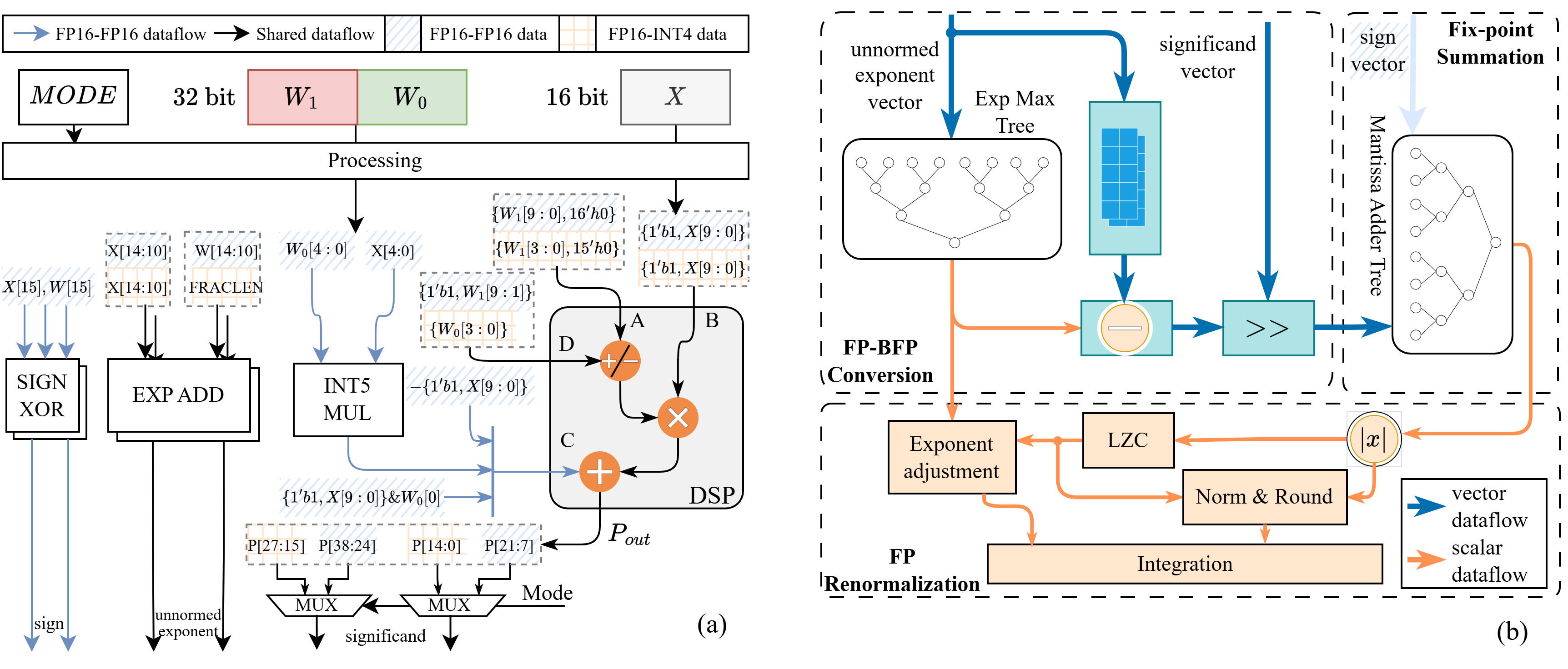}
    \caption{(a) The detailed design of mixed precision PE. (b) FP16 accumulation tree with BFP-FP conversion}
    \label{fig:PE}\vspace{-1.5em}
\end{figure}

To mitigate this overhead, our key insight is to \emph{decouple accumulation from full floating-point representation and to perform accumulation within a unified numerical domain.} 
Specifically, as shown in \Cref{fig:PE}.(b), instead of materializing FP16 values after each DSP operation, our accumulation tree
\revC{collects a set of unnormalized partial results (\textit{i.e.} significand product from DSP and unbiased exponent result from addition) from the same PE array column at first.
Then all exponent results participate in a maximum-exponent reduction tree to determine the global maximum $E_{max}$. 
Each significand is then arithmetic right-shifted by $E_i - E_{max}$ to be aligned to the same binary point position. 
After alignment, the shifted significands are accumulated directly using a conventional fixed-point adder tree, producing a BFP representation composed of a shared exponent and an aggregated fixed-point significand.
Before converting the accumulated result back to FP16 format, a renormalization stage is performed to restore the standard normalized floating-point representation, $1.\text{mantissa} \times 2 ^{E-\text{bias}}$. 
A leading-zero counter (LZC) identifies the most significant non-zero bit of the accumulated significand, after which the significand is shifted accordingly and the exponent is adjusted to compensate for the normalization offset. 
Finally, the normalized result is truncated to fit the FP16 mantissa width.
}
This design substantially simplifies the accumulation data path and also enables a lower-cost mixed-precision PE architecture, which is shown in \Cref{fig:PE}.(a). 

By delegating the floating-point data conversion to the accumulation tree, the PE micro-architecture is streamlined to primarily consist of the DSP48E2 unit for mantissa multiplications, an auxiliary 5-bit multiplier for corruption recovery and two INT adders to produce the unnormalized exponents. 
In FP16-FP16 mode, the PE follows overpacking scheme by first producing corrupted significands from two concurrent FP16 multiplications and then injecting the truncated information concatenated with the result of the INT5 multiplier to recover clean intermediate fractions. 
The sign of each mantissa product is determined via a simple XOR operation and propagated to the mantissa accumulation tree, which uses this sign information to correctly accumulate the absolute values produced. 
\revC{
To further reduce the hardware cost of the accumulation tree, we truncate the bit-width of the significand inputs and retain only the 15 MSBs rather than full 22-bit products. 
This design choice is motived by the limited effective precision of the target FP16 format. 
Specifically, FP16 arithmetic preserves only an 11-bit effective significand precision (including the hidden bit) after normalization, meaning that a large portion of the lower-order multiplication bits ultimately contributes only to rounding behavior rather than the final numerical representation.
To achieve better accumulation accuracy, we retain several additional guard bits beyond the nominal FP16 significand precision. 
In particular, the 15-bit truncation bit width is selected to account for carry propagation effects introduced during multi-operand accumulation, where lower-order bits may still influence higher-order significand regions.
The experimental results presented in \Cref{sec:pe_validation} further validate this design choice, where the proposed 15-bit truncation consistently outperforms standard cascaded FP16 accumulation across all evaluated workloads. 
}
In FP16-INT4 mode, the standard DSP packing is adopted without the introduction of overpacking and corruption recovery. 
The INT4 data in the 2's complementary format are directly fed into the DSP ports and the products of the DSP units can be used in the mantissa accumulation tree without the extra sign information. 

\subsubsection{Validation} 
\label{sec:pe_validation}
To validate the efficiency of our proposed scheme, we design additional sets of control experiments. 
In these experiments, 64 cascaded FP MAC IPs and overpacked DSPs are used as the baseline while several variants of our PE array are implemented to explore the trade-off between resource utilization and numerical fidelity. 
Specifically, IMPL1 and IMPL2 correspond to computation engines whose BFP accumulation trees are implemented using LUT-based adders that accept 22-bit and 15-bit operands, respectively.
Notably, the summation of 64 signed 15-bit values produces at most a 22-bit result while each DSP adder supports up to 48-bit addition, enabling two 15-bit accumulation paths to be packed into a single DSP.
Leveraging this property, the LUT consumption of the accumulation tree can be significantly reduced by implementing it using DSP resources, resulting in IMPL3. 
The numerical fidelity is evaluated under two input settings: randomized data and data sampled from the empirical distribution of weights and activations in the Llama-2 model. 
As can be seen in \Cref{tab:PE_COMPARE}, our PE array design can save approximately $57.2\%$ LUT resources under the same DSP consumption and can always achieve better computation accuracy in random and empirical settings. 

\renewcommand{\arraystretch}{0.2}
\begin{table}[htp]
\centering
\caption{Performance comparison of the mixed-precision computation unit.}
    \begin{threeparttable}
    \begin{tabular}{c c c c c c c c}
    \toprule
    \multirow{9}{*}{Name} & \multicolumn{4}{c}{Computation Error($\%$)} & \multirow{9}{*}{LUT$^*$} &  \multirow{9}{*}{FF$^*$} &  \multirow{9}{*}{DSP$^*$} \\
    \cmidrule{2-5}
    & \multicolumn{2}{c}{Random} & \multicolumn{2}{c}{Empiricial} & & & \\
    \cmidrule{2-3} \cmidrule{4-5}
    & FP16$\times$FP16 & FP16$\times$INT4$^\dagger$ & FP16$\times$FP16 & FP16$\times$INT4$^\dagger$ & & & \\
    \midrule
    IMPL1$^\ddagger$ & 0.035 & 0.035 & 0.065 & 0.036 & 1544+6829 & 3618+6043 & 32 \\
    IMPL2$^\ddagger$ & 0.064 & 0.074 & 0.116 & 0.067 & 1472+5244 & 3426+4850 & 32 \\
    IMPL3$^\ddagger$ & 0.064 & 0.074 & 0.116 & 0.067 & 1408+4408 & 3426+4138 & 32 + 31.5 \\
    \cmidrule{1-8}
    Cascade MAC IP & 0.410 & 0.369 & 0.490 & 0.415 & 13598 & 21138 & 64 \\
    \cmidrule{1-8}
    Cascade Overpacked DSP & 0.390 & 0.370 & 0.344 & 0.348 & 9920 & 7264 & 32 \\
    \bottomrule
    \end{tabular}
    \begin{tablenotes}
        \small
        \item[$^*$] resource consumption is averaged for one column of accumulation.
        \item[$^\dagger$] computation error is reported before dequantization for FP16$\times$INT4 case
        \revC{\item[$^\ddagger$] The resource consumption is reported as "PEs + accumulation tree"}
    \end{tablenotes}
    \end{threeparttable}
    \label{tab:PE_COMPARE}
\end{table}

\subsection{Tile-based nonlinear processing engine}

\begin{figure}[htbp]
    \centering
    \vspace{0cm}
    \includegraphics[width=0.95\textwidth]{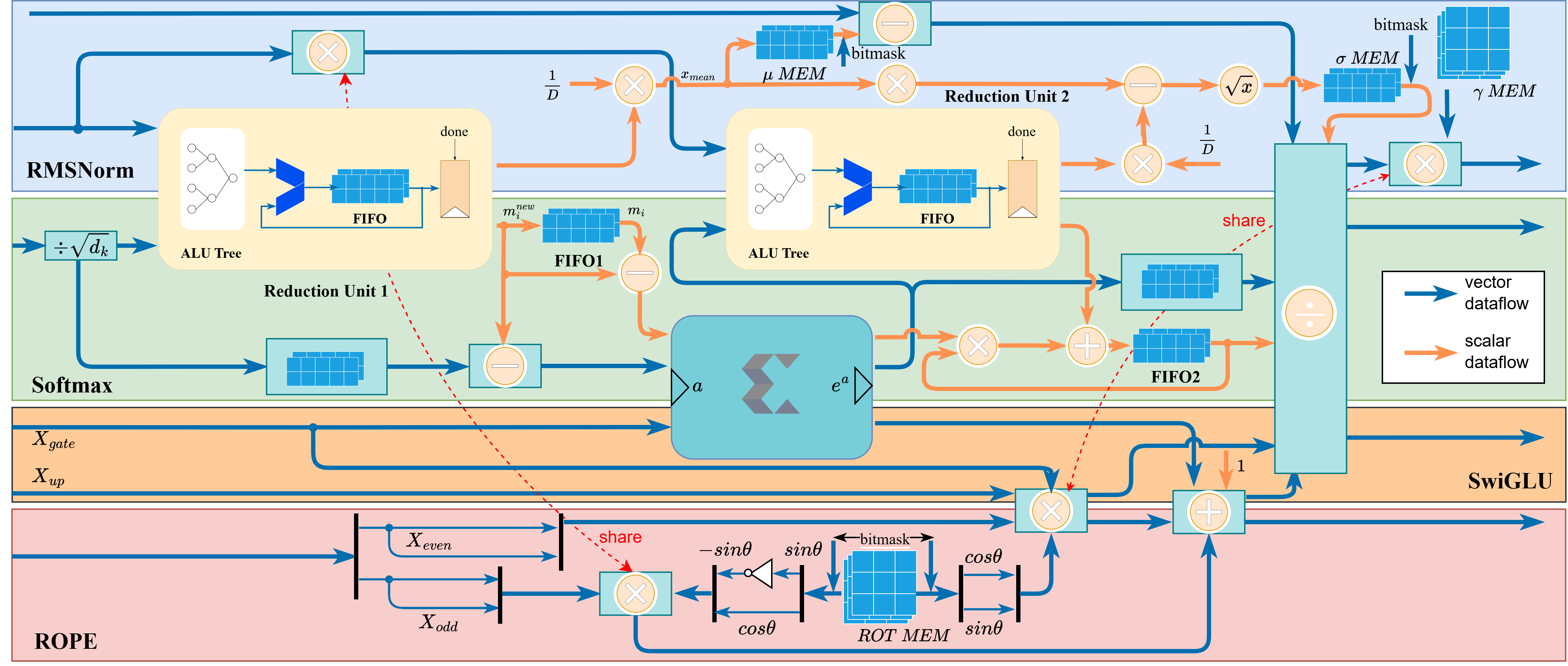}
    \caption{The micro-architecture of nonlinear processing engine(NPE).}
    \label{fig:NPE}\vspace{-1.5em}
\end{figure}

\Cref{fig:NPE} illustrates the micro-architecture of the proposed nonlinear processing engine (NPE), 
\revC{
which inherits the throughput-matching and resource-sharing design principles of METAL~\cite{METAL}. 
To achieve a unified and resource-efficient execution dataflow, the NPE takes advantage of a key scheduling insight: both the FlashAttention-style Softmax (requiring running maximums and exponential sums) and the RMSNorm modules (requiring mean and variance) demands two concurrent reduction operations during their respective calculation loops, yet their executions are completely non-overlapping, ensuring only one non-linear function is active at a time.
Utilizing this mutual exclusivity, both non-linear submodules route their operations through the shared Reduction Units. To support the different reduction functionalities of each dataflow, this unit incorporates a reconfigurable reduction tree composed of ALUs that can dynamically switch between maximum ($max$) and addition ($sum$) operations.
The computation within this unified dataflow operates in a non-blocking, two-phase paradigm. 
In the first phase, the NPE accepts output matrix tiles row-by-row. 
Activations pass through the pipelined computation units, and intermediate reduction features are stored in dedicated row-wise intermediate FIFOs. 
In successive computational rounds of the same row, these partial features are popped in-order and updated incrementally until the entire row is traversed, at which point the finalized numerical features are written back to the on-chip buffers. 
In the second phase, the original matrix tiles are streamed into the NPE again without triggering feature updates; 
instead, the stored features are accessed sequentially to finish the normalization process.  
}
Because the numerical features for RMSNorm are computed before execution decisions are finalized, a bitmask mechanism is introduced to guide the RMSNorm unit in selectively reading only the features corresponding to unskipped tokens.

Although SwiGLU and RoPE do not involve reduction operations, their dataflows are naturally integrated into the NPE pipeline with minimal structural overhead. 
For SwiGLU, the computation is formulated as $SILU(XW_{gate}) \odot XW_{up}$, requiring support for the $SiLU$ activation and a Hadamard product between two matrices. 
The $SiLU$ can be simply implemented by reusing the existing exponential unit and division unit while the Hadamard product can be realized through vector-wise multiplications on a cycle-by-cycle basis, leveraging its element-wise nature.
To simplify data orchestration, $W_{up}$ and $W_{gate}$ are interleaved in off-chip memory, enabling the PE array to generate paired output tiles from $XW_{up}$ and $XW_{gate}$ in each computation round on the fly. 
These tiles are then consumed synchronously by the NPE to generate output tiles of matching shape.


For RoPE, complex-domain rotation is implemented through pair-wise transformations by partitioning each token vector into even-indexed and odd-indexed elements, with identical reorganization applied to the rotary embedding parameters. 
The required multiplications and additions between the reorganized vectors are realized by further reusing the multiplication units, producing the correct rotated representations. 
The rotary embedding tensors are precomputed on the CPU host and stored in off-chip memory. 
\revC{
A dedicated ROT (Rotary) buffer implemented via standard on-chip BRAMs acts as a localized memory to store those pre-computed rotary embedding parameters (cosine and sine coefficients) during execution. This allows the specialized RoPE submodule to perform fine-grained phase rotations on incoming token vectors without incurring off-chip memory access.
}
Since token indices subject to RoPE are non-contiguous due to dynamic pruning, a bitmask mechanism is again employed to control selective reading the embedding parameters stored in ROT buffer for rotary transformation application.

\subsection{Optimization with cross-layer KV reuse}
\label{sec:memory_opt}
\subsubsection{KV memory mapping} 
To fully exploit the bandwidth of HBM devices through AXI interface, model parameters and KV caches must be distributed across channels in a carefully designed memory mapping scheme that enables long burst transactions and maximizes channel-level parallelism. 
In conventional LLM inference, KV entries are generated and consumed in a strictly sequential manner along the token dimension, making it effective to interleave KV tensors across HBM channels. 
Such an interleaved layout aligns consecutive KV accesses with contiguous memory regions, thereby preserving AXI burst locality while exploiting the inherent parallelism provided by multiple channels.

However, this fact no longer holds under dynamic pruning with KV reuse. 
With token-wise layer skipping, only a subset of tokens produces new KV entries at a given layer, while missing entries are substituted with KV values reused from earlier layers to preserve attention semantics. 
This mismatch between KV generation and KV consumption leads to highly irregular access patterns during attention computation: KV entries corresponding to skipped tokens are fetched from different layers and non-contiguous memory locations. 
As a result, AXI burst transactions are frequently fragmented, severely degrading effective HBM bandwidth utilization.

\begin{figure}
    \centering
    \vspace{0cm}
    \includegraphics[width=0.95\textwidth]{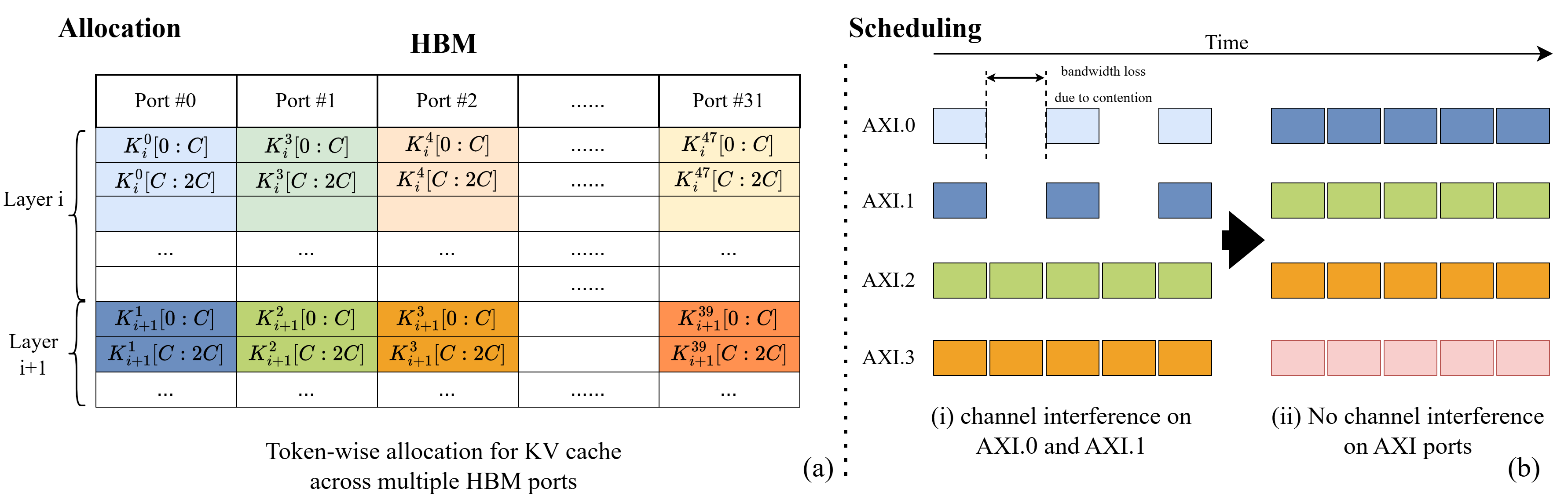}
    \caption{(a) Token-wise memory mapping scheme for KV cache (b) Channel interference to load KV cache from previous layer.}
    \label{fig:mapping}\vspace{-1.5em}
\end{figure}

To address this challenge, we propose a token-wise memory mapping scheme tailored to the pruning granularity of SkipGPT-style models. 
As illustrated in \Cref{fig:mapping}.(a), the KV entry of a given unskipped token is placed entirely within the local address space of a single port (pseudo-channel) on HBM AXI interface, rather than being interleaved between ports. 
And the KV caches of the unskipped tokens are distributed across HBM ports in a round-robin manner, allowing KV segments of multiple tokens to be fetched concurrently across ports. 
This layout effectively aligns the token-level parallelism of attention computation with the port-level parallelism of HBM. 
Moreover, KV accesses for each token can be served through long, contiguous burst reads within a port, regardless of whether the KV entries originate from the current layer or are reused from previous layers. 
Specifically, the address span required per token can be calculated as $ D\times DW/HBM\_PORT\_WIDTH$, which is sufficient to sustain large burst transactions. 

\subsubsection{KV invariance buffer} 
The token-wise memory mapping scheme could still introduce channel interference under certain access patterns, as illustrated in \Cref{fig:mapping}.(b). 
During decoding at layer $i+1$, self-attention for an unskipped token requires not only the newly generated KV entries at layer $i+1$, but also KV entries of skipped tokens reused from earlier layers to preserve attention semantics. 
For example, fetching $K_{i+1}^1$ and $K_{i}^0$ concurrently may target the same physical HBM channel due to token-wise port assignment, while other HBM physical channels (e.g.,  physical channel associated with Port $\#3$ in this case) remain idle.
As a result, simultaneous AXI transactions contend for the same channel, degrading memory throughput despite sufficient interface-level parallelism.

To mitigate this issue, we pin reused KV entries in an on-chip invariance buffer. 
By serving reused data directly from the on-chip buffer, off-chip HBM accesses are restricted exclusively to KV entries generated at the current layer. Under this condition, each HBM port fetches data only from its locally assigned memory region, and no cross-port physical channel conflicts are introduced. 
As a result, port-level parallelism directly translates into independent HBM physical channel activity, ensuring balanced channel utilization and maximizing effective bandwidth, as illustrated in \Cref{fig:mapping}(b)(ii).

We argue that managing this buffer \revC{incurs negligible performance overhead} with pre-determined routing decisions.
At the start of the decoding of token $j$ at layer $i$, the routing bitmask for layer $i+1$ regarding all preceding tokens is already finalized during the generation of token $j-1$. This look-ahead information allows the controller to determine—in the background—which KV entries (whether currently being fetched from HBM or residing in the buffer) must be retained for the subsequent layer. 
Consequently, the invariant KV selection and the related buffer update for layer $i+1$ is performed concurrently with the decoding of layer $i$. 
\revC{
Specifically, the data fetched from the HBM interface is broadcast to both the PE array for linear projections ($QK^T$ and $SV$ computations) and the KV buffer update logic. 
A crossbar-based routing mechanism selects and stores entries required for subsequent self-attention computation.
This design enables buffer maintenance to be fully overlapped with computation and to introduce no additional off-chip memory transactions, thereby removing it from the critical path of AXI data movement.
}

This \revC{overlap-based latency hiding} provides maximum benefit during consecutive self-attention layers, where the buffer acts as the primary data source for reused KVs. 
In scenarios where self-attention is skipped at layer $i+1$, the invariance buffer is effectively invalidated. 
When self-attention eventually resumes at a later layer, cross-layer KV loads from HBM are required to restore attention correctness. However, even in this "non-consecutive" case, the proposed approach does not degrade performance; the buffer update for the next potential attention process remains an auxiliary, overlapped operation that introduces no additional contention compared to the baseline execution.
By significantly accelerating consecutive-layer blocks—the most common pattern in our slightly pruned LLMs—we achieve a substantial increase in effective memory throughput.

\subsubsection{KV Cache Scheduling}
\begin{figure}
    \centering
    \vspace{0cm}
    \includegraphics[width=0.95\textwidth]{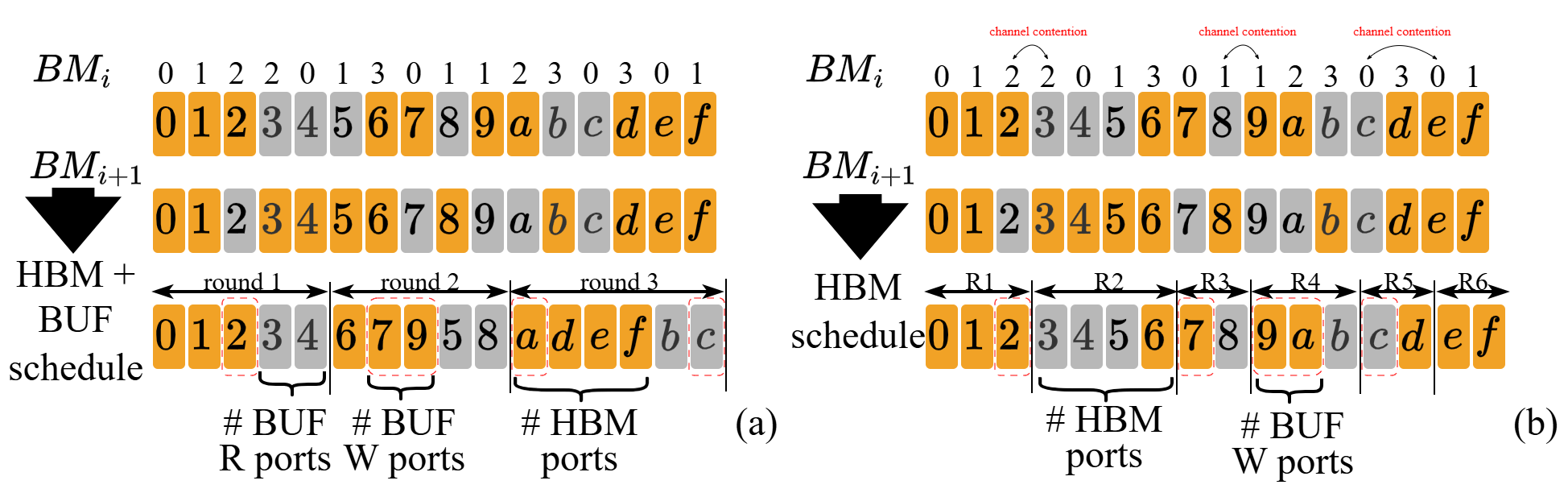}
    \caption{(a) Valid KV cache scheduling. (b) Invalid KV cache scheduling.}
    \label{fig:cache}\vspace{-1.5em}
\end{figure}

\Cref{fig:cache} illustrates the cycle-accurate scheduling of KV cache access patterns under the proposed invariance buffer management system. The bitmasks ($BM_i$, $BM_{i+1}$) represent the routing decisions for sixteen tokens (index $0-f$) at layer $i$ and layer $i+1$, respectively, where orange blocks denote active attention tokens and gray blocks represent skipped tokens whose KV entries are reused.
For this visualization, we assume a hardware configuration with four HBM ports and two invariance buffer ports. 
The numerical indices above $BM_i$ designate the specific HBM port to which each token is mapped. 
Notably, while unskipped tokens are assigned to HBM ports in a deterministic round-robin manner to maximize spatial parallelism, skipped tokens exhibit no such pattern, as they are sourced from disparate preceding layers.

\textbf{Case-1: Consecutive-Layer Attention} \Cref{fig:cache}.(a) demonstrates the scheduling when the invariance buffer is valid for decoding at layer $i$.
By serving reused (gray) entries via the invariance buffer, the HBM interface is dedicated exclusively to the KV (orange) entries at the current layer. 
The scheduling width per round is governed by the structural port constraints of both the HBM and the buffer. 
For instance, in Round 1, the schedule is limited to two reused entries to match the two-port buffer read capacity; In Round 2, the schedule is capped at three unskipped entries.
This ensures that the concurrent proactive updates for layer $i+1$ (red dashed boxes) do not exceed the buffer’s write-port capacity, thereby maintaining the zero-overhead property.

\textbf{Case-2: Non-Consecutive Layer Attention} \Cref{fig:cache}.(b) depicts the scenario where the self-attention at layer $i-1$ is skipped. Under this case, the invariance buffer has been evicted prior to decoding at layer $i$, forcing the accelerator to read cross-layer KV caches from HBM. 
The lack of a structured pattern among reused tokens leads to frequent HBM port collisions. 
To prevent channel contention and maintain throughput symmetry across AXI ports, the scheduling width per round is primarily restricted to ensure no channel interference occurs between concurrent tokens.
The resulting serialized, lower-throughput schedule highlights how the invariance buffer effectively transforms unpatterned HBM traffic into high-bandwidth, deterministic on-chip accesses.

\subsubsection{Permutation-Invariant Processing}
To further minimize FPGA resource consumption, our design exploits the permutation invariance of the attention mechanism. 
Rather than reordering the locally permuted tokens within each fetch back into their original sequence, the concatenated data are streamed directly into the PE array for $QK^T$ and $SV$ computation. 
Although the out-of-order $QK^T$ stage produces a score vector with permuted local indices, the final attention reduction is a sum-based operation, making the result numerically invariant to the arrival sequence.
Moreover, as the routing bitmasks ($BM_i$, $BM_{i+1}$) for Key ($K$) and Value ($V$) entries are identical, the elements of the resulting score vector and their corresponding $V$ entries remain structurally coupled across the dataflow.
Consequently, the subsequent value-weighting ($SV$) stage can operate on these permuted pairs to produce the correct self-attention output. This design choice eliminates the need for high-area Reorder Buffers (ROB) or complex crossbar switches to restore a strict sequential index, significantly reducing LUT and FF utilization.

\section{Experiments}
\subsection{Experiment setup}
\subsubsection{Workload} We evaluate \textbf{SkipOPU} with Llama2\cite{Llama2} and apply SkipGPT methodology to prune Llama2 models with around 25\% skipping probability. We then symmetrically quantize the model parameters to 4-bit fixed-point representation using GTPQ\cite{GPTQ} while the $KV$ cache and intermediate activation matrix remain FP16 format. 
The workload includes prefill stage with context in the length of 128, 256, 512, 1024 and decode stage with 512, 1024 output tokens. We use batch size 1 for evaluation as it is the most common case in the edge application.

\subsubsection{Implementation} 
\revC{
\textbf{SkipOPU} is written entirely in fully synthesizable SystemVerilog adhering to the IEEE 1800-2017 standard, and is synthesized and implemented utilizing Vivado 2020.1 targeting an AMD Alveo U280 acceleration platform. Handcrafted RTL development was deliberately selected over HLS to maintain fine-grained control over our dense logic fabric, prevent unpredictable resource inflation and enforce strict post-routing timing closure. 
More importantly, explicit RTL-level design is mandatory to support our specialized multi-precision arithmetic configurations; only by directly instantiating and controlling the underlying DSP48E2 units through direct primitive-level description can the control logic manipulate dynamic hardware signals to support mixed-precision computation at runtime. 

The accelerator core operates at a base frequency of 225MHz, while the HBM2 interface runs at 450MHZ according to the architectural justification discussed in \Cref{sec:overview}. System integration incorporates standard infrastructure IP blocks, including the AMD/Xilinx HBM2 Controller, the MIG for DDR4 SDRAM, PCIe host interfaces, and AXI SmartConnect blocks. These infrastructure subsystems were pre-configured and wrapped using the Vivado IP Catalog, and modularly instantiated by module name within the top-level fabric. Functional verification was conducted through testbench simulations via Modelsim 2019.4 SE, while correctness was ensured by comparison against the golden case extracted from PyTorch baseline. 
}

\revC{
\subsubsection{Memory mapping}
\begin{table}[htbp]
\caption{Heterogeneous memory subsystem specifications on Alveo U280}
\label{tab:mem_summary}
\centering
\small
\begin{tabular}{cccccc}
\toprule
\textbf{Storage Tier} & \textbf{Hardware Primitives} & \textbf{Bandwidth} & \textbf{Memory capacity} & \textbf{AXI Configuration/Bus Width} \\ 
\midrule
\multirow{2}{*}{Off-Chip} & HBM2 & \multirow{2}{*}{460.8 GB/s} & \multirow{2}{*}{8GB} & 256-bit, Dynamic Burst \\
& (32 pseudo-channels) & & & (Max BL = 256) \\ \addlinespace
\multirow{2}{*}{Off-Chip} & DDR4 & \multirow{2}{*}{38 GB/s} & \multirow{2}{*}{32GB} &512-bit, Dynamic Burst \\ 
& (Dual-channel SDRAM) & & & (Max BL = 256) \\ \addlinespace
On-Chip & BRAM & Design-dependent & 7.56MB & Configurable (Up to 72-bit) \\ \addlinespace
On-Chip & URAM & Design-dependent & 33.75MB & fixed 72-bit \\
\bottomrule
\end{tabular}
\end{table}
To maximize PE utilization and eliminate memory access bottlenecks across the heterogeneous storage hierarchy of the Alveo U280, \textbf{SkipOPU} implements a deterministic memory allocation strategy. 
This memory mapping prevents the ultra-high-bandwidth HBM channels from being throttled by small metadata fetches while utilizing on-chip buffer resources to improve data reuse across operations and hide off-chip memory access latency. 
The hardware specifications of these underlying storage primitives are summarized in \Cref{tab:mem_summary}. 

The memory allocation strategy is governed by the distinct bandwidth and latency profiles of each storage tier. 
Specifically, all weight matrices and active KV cache tensors are pinned exclusively to HBM2, leveraging its ultra-high bandwidth to maximize throughput during the decoding phase. 
Conversely, the activation matrices and small volume parameters (\textit{e.g.}, RoPE constants, RMSNorm coefficients, and router metadata) are stored in DDR4, whose access latency can be well hidden by exploiting data reuse via the on-chip caching resources. 
The on-chip localized memory hierarchy is further partitioned based on storage density and parallel computation requirements. 
High-density UltraRAM (URAM) blocks are utilized to cache sustained KV cache tensors between successive layers to eliminate redundant off-chip traffic. 
Meanwhile, Block RAMs (BRAMs) are dedicated to storing active data chunks during computation, providing highly configurable, wide-bus access capabilities that satisfy processing parallelism and eliminate pipeline stalls.
}

\subsection{Resource breakdown}
\Cref{tab:resource} summarizes the resource utilization of \textbf{SkipOPU}. The processing-element (PE) array integrates 4096 DSPs and, enabled by the fine-grained packing technique described earlier, sustains $64\times128$ FP16$\times$FP16 or FP16$\times$INT4 multiplications per cycle.
\revC{
Crucially, our overpacked PE array implementation is strictly required to render the entire core synthesizable and routable within the platform's physical boundaries. 
As demonstrated by the control experiments evaluated in \Cref{tab:PE_COMPARE}(IMPL2 vs. Cascaded MAC IPs), an unpacked implementation requires 13,598 LUTs per accumulation column, compared to only 6716 (1472+5244) LUTs for our overpacked design. 
Extrapolating the unpacked configuration across our full 128-column PE array yields a projected demand of $1,740,544$ LUTs. This exceeds the total physical logic capacity of the AMD Alveo U280 platform ($1,303,680$ available LUTs) by over 33\%, leading to severe resource exhaustion and routing failure. 
Consequently, the overpacking technique serves as the critical design choice to compress the massive arithmetic logic footprint, explaining why the design achieves a dense global LUT utilization while keeping DSP consumption bounded at $51\%$.
}

To match the throughput of the PE array, a 64-way tile-wise nonlinear processing engine (NPE) is introduced, incurring an additional 16.78\% LUT and 2.95\% DSP overhead to support miscellaneous nonlinear operations. 
The DMA (Direct Memory Access) engine, which interfaces with both DDR4 and HBM for data movement, consumes 3.52\% LUT, 2.26\% FF, and 20.00\% BRAM resources.
To enable the KV invariance buffer, 512 URAMs, capable of storing up to 1024 tokens' KV entries, are organized into a memory interface with 16 independent ports, each of which is designed to match the throughput of the corresponding HBM ports. Concurrent KV updates during attention computation are supported by a $48\times16$ crossbar that dynamically selects reusable KV values. Overall, this memory subsystem (KV BUF + XBar) accounts for 8.18\% LUT, 0.57\% FF, and 53.33\% URAM utilization.

\renewcommand{\arraystretch}{1.0}
\begin{table}[htp]
    \centering
    \caption{Skip-OPU resource breakdown on U280 FPGA}
    \begin{tabular}{c|ccccc}
        \hline
        component & LUT & FF & DSP & BRAM & URAM \\
        \hline
        PE array & 853184 (65.44\%) & 1085976 (41.65\%) & 4096 (45.39\%) & 0 (0.0\%) & 0 (0.0\%) \\
        Tile-wise NPE & 218830(16.78\%) & 109835 (4.21\%) & 266 (2.95\%) & 29 (1.44\%) & 0 (0.0\%) \\
        DMA Engine & 45947 (3.52\%) & 58814 (2.26\%) & 0 (0.0\%) & 383 (20.00\%) & 0 (0.0\%)\\
        KV BUF +Xbar& 106704 (8.18\%) & 14928 (0.57\%) & 0 (0.0\%) & 0 (0.0\%) & 512 (53.33\%)\\
        Total & 1288673(98.85\%) & 1668477(63.99\%) & 4618 (51.17\%) & 775 (38.44\%) & 512 (53.33\%) \\
        \hline
    \end{tabular}
    \label{tab:resource}
\end{table}

\revA{
\subsection{Model accuracy evaluation}
We evaluate the model perplexity of the Llama2 family across the WikiText2\cite{WT2} and PTB\cite{PTB} datasets under different compression methods in \Cref{tab:perplexity_wide}. 
We take the original dense model with FP16 precision as the baseline, and SkipOPU includes both INT4 weight quantization and dynamic layer pruning. 
To isolate the exact sources of quality variation, we also compare standalone dynamic layer skipping without KV cache reuse (SkipLayer~\cite{LearnToSkip}), dynamic skipping with KV reuse (SkipGPT~\cite{SkipGPT}), and INT4 weight quantization (AWQ~\cite{AWQ}). 
Compared with the baseline, the perplexity of SkipOPU experiences a predictable, compounding increase (\textit{e.g.}, from 5.47 to 8.52 for Llama2-7B on WikiText2). 
However, the resulting perplexity of SkipOPU remains closely comparable to standalone dynamic skipping methods like SkipLayer and SkipGPT (\textit{e.g.} 5.67 vs 6.78 vs 7.84). 
This indicates that the concurrent inclusion of W4 quantization and KV cache reuse introduces an acceptable and expected level of algorithmic degradation. 
These evaluations strongly validate our stacked compression choices, demonstrating that competitive model quality is preserved while establishing a favorable design space for our architectural optimizations.

\begin{table}[t]
\centering
\caption{The model perplexity across different datasets under comprehensive compression methods.}
\label{tab:perplexity_wide}
\begin{tabular}{@{}ccccccc@{}}
\toprule
\multirow{2}{*}{\textbf{LLMs}} & \multirow{2}{*}{\textbf{Dataset}} & \multicolumn{5}{c}{\textbf{Compression Methods}} \\
\cmidrule(l){3-7} 
& & \textbf{Baseline} & \textbf{SkipLayer} & \textbf{SkipGPT} & \textbf{AWQ} & \textbf{SkipOPU (Ours)} \\ 
\midrule
\multirow{2}{*}{Llama2-7B}  & WikiText2 & 5.47 & 5.86 & 7.81 & 5.54 & 8.52 \\
                            & PTB & 20.83 & 28.03 & 30.58 & 24.14 & 35.96 \\ \midrule
\multirow{2}{*}{Llama2-13B} & WikiText2 & 4.88 & 5.67 & 6.78 & 5.96 & 7.84 \\
                            & PTB & 28.92 & 35.84 & 41.69 & 30.59 & 45.51 \\ 
\bottomrule
\end{tabular}
\end{table}
}
\subsection{Impact of dataflow optimization}
\begin{figure}
    \centering
    \includegraphics[width=0.9\textwidth]{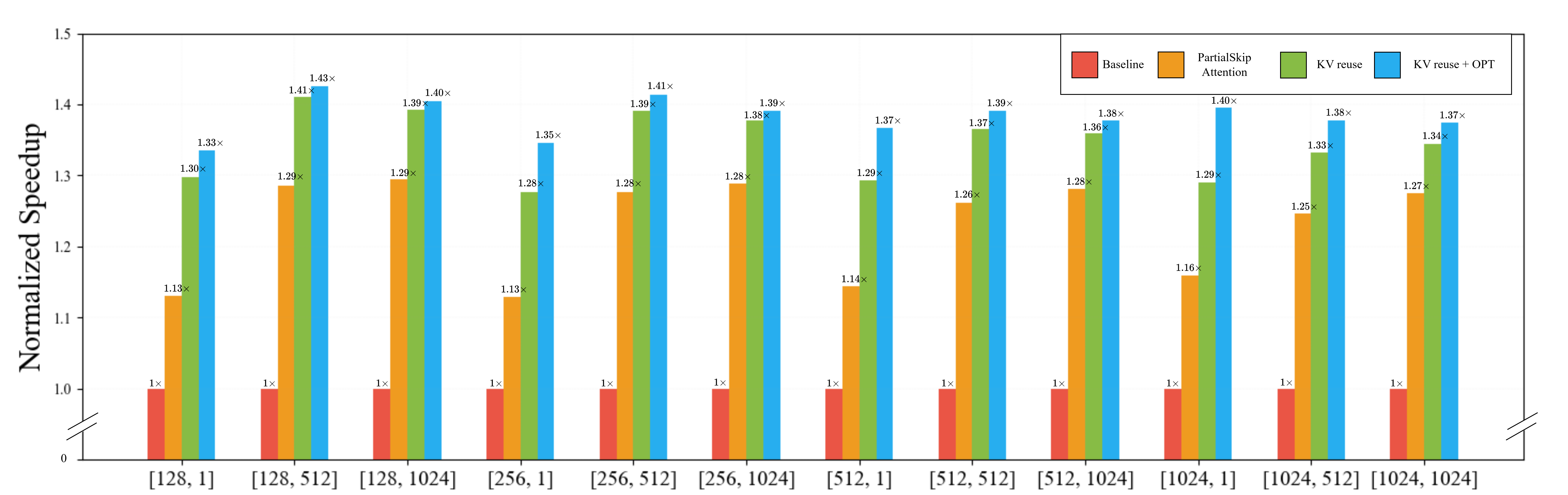}
    \caption{Normalized MHA speedup under different dataflow optimization and varying [prefill:decode]}
    \label{fig:speedup}\vspace{-1.5em}
\end{figure}

We first focus on the computation of the MHA submodule that includes the router, RMSNorm and softmax to evaluate the performance impact of the proposed fused dataflow under the previously described workloads. 
Specifically, four progressively enhanced configurations are evaluated, each introducing an additional optimization on top of the previous one.
\begin{itemize}
    \item \emph{Baseline:} This configuration strictly follows the original Llama-2 computation graph, where all tokens across all layers execute identical operations. Nonlinear operations are implemented using a conventional row-wise dataflow, incurring inherent buffering stalls between linear and nonlinear stages.  
    \item \emph{PartialSkip Attention:} \revC{It has the same configuration as \emph{Baseline} except} a linear router is inserted before the RMSNorm operation to determine whether the corresponding MHA computation should be executed or skipped on a per token-basis. To preserve attention semantics, KV caches are still generated for skipped tokens, while only the attention computation itself is conditionally bypassed. 
    \item \emph{KV reuse:} The same routing mechanism is retained, but KV caches for skipped tokens are no longer recomputed;
    instead, KV values from earlier layers are reused for missing entries. \revC{In this case, row-wise computation pattern is still used for nonlinear operations.}
    \item \emph{KV reuse $+$ OPT}: The proposed dataflow optimizations are applied on top of KV reuse, including nonlinear latency hiding and multi-head packing in the fused attention dataflow.
\end{itemize}
The performance of \textbf{SkipOPU} with incremental dataflow optimizations, where we consistently apply token-level memory mapping scheme to reused KV cache for fair comparison, is shown in \Cref{fig:speedup}.
At the prefill stage, the improvement of \emph{PartialSkip} and \emph{KV Reuse} stems from the reduced arithmetic complexity and remains relatively stable across sequence lengths, around $1.14\times$ and $1.29\times$,  respectively. 
\revC{
The fused dataflow optimization further improves the MHA throughput to $1.40\times$ by hiding the nonlinear operation latency via tile-wise incremental computation, which eliminates the pipeline bubble caused by waiting the full rows of resulting matrix to begin execution.
}

When it comes to the decoding phase, the system transitions to a memory-bound regime. 
Here, the primary benefit of attention skipping arises from the reduction in memory transactions for model parameters and KV entries. 
While the latency of nonlinear operations on a single vector is negligible in this phase, the speedup from the fused dataflow is gradually diminished as the decoding sequence length grows.
A critical observation is the diminishing performance gap between \emph{PartialSkip} and \emph{KV Reuse} as the context length expands. 
This phenomenon is driven by the fact that cross-layer KV retrieval introduces persistent channel contention, even when employing a token-level memory mapping scheme. 
As the volume of retrieved KV entries grows, the associated memory traffic begins to dominate the total execution cost. 
The resulting degradation in effective memory bandwidth for KV loads effectively offsets the reduction in total memory transactions, highlighting the necessity of our KV Invariance Buffer to transform unpatterned
HBM traffic into deterministic on-chip accesses.

\subsection{Impact of allocation and scheduling}
Building upon our $KV$ cache reuse and fused dataflow optimizations, we analyzed the impact of our token-level memory mapping scheme and the corresponding $KV$ invariance buffer on effective HBM bandwidth.
\revC{To quantify the bandwidth lost incurred by algorithmic sparsity and demonstrate the effectiveness of our proposed hardware optimizations, we define four progressive ablation settings:}

\revC{
\begin{itemize}
    \item \textit{Dense (Baseline):} Follows the original Llama-2 computation graph, where every token is processed equivalently and no sparsity is applied.
    \item \textit{Sparse + Int.:} Introduces dynamic layer skipping and $KV$ reuse during computation, while retaining a conventional interleaved memory mapping scheme where the $KV$ entries of individual tokens are interleaved across multiple HBM channels.
    \item \textit{Sparse + Token:} Utilizes token-wise memory mapping to allocate the fragmented active $KV$ cache under the $KV$ reuse strategy.
    \item \textit{SkipOPU (Full System):} Further instantiates the $KV$ invariance buffer on-chip to sustain $KV$ entries that will be reused across adjacent layers. 
\end{itemize}
}
Our baseline evaluation focuses on the \textit{Dense} case. In this state, the naturally contiguous memory footprint allows the HBM controller to maintain highly efficient, long-burst transactions, achieving an effective bandwidth of 408.7 GB/s (88.7\% of peak). 
As we move toward the more memory-efficient \textit{Sparse + Int.} scenario utilizing $KV$ reuse, the total data volume is reduced by approximately $25\%$, which significantly lowers the HBM capacity requirement for long-context generation. However, this reduction in volume comes at the cost of a fragmented access pattern: under the standard interleaved mapping scheme, the controller suffers from severe row-buffer thrashing as it attempts to switch between different memory regions across layers. This ``context-switching'' tax creates severe Page Miss penalties, collapsing the effective HBM bandwidth to a worst-case performance of only 55.8\%.
To reclaim this lost throughput in the \textit{Sparse + Token} setting, we implemented a token-level memory mapping scheme that restores spatial locality by pinning $KV$ entries to specific ports, ensuring $KV$ accesses for each token are served through long, contiguous burst reads. While this mapping introduces minor inter-channel contention during concurrent reads, it successfully lifts the peak performance back to 360.2 GB/s.
\revC{
Finally, we introduce our $KV$ invariance buffer in the \textit{SkipOPU} case, which proactively caches reused entries in on-chip URAM during the attention phase of previous layers. By serving repeated values directly from high-speed URAM, we reduce not only redundant HBM reads but also the penalties associated with cross-layer channel interference. Even though the buffer cannot eliminate cross-layer $KV$ reads in the case of non-consecutive self-attention (which limits raw physical HBM utilization to $83.1\%$), the synergy between the on-chip buffer and the HBM interface enables an aggregate effective bandwidth of 467.8 GB/s. 
This result highlights that \emph{SkipOPU} achieves near-saturation of the HBM bandwidth while further amplifying effective memory throughput through hierarchical reuse with the introduction of the $KV$ invariance buffer.
}
\begin{figure}
    \centering
    \includegraphics[width=0.9\textwidth]{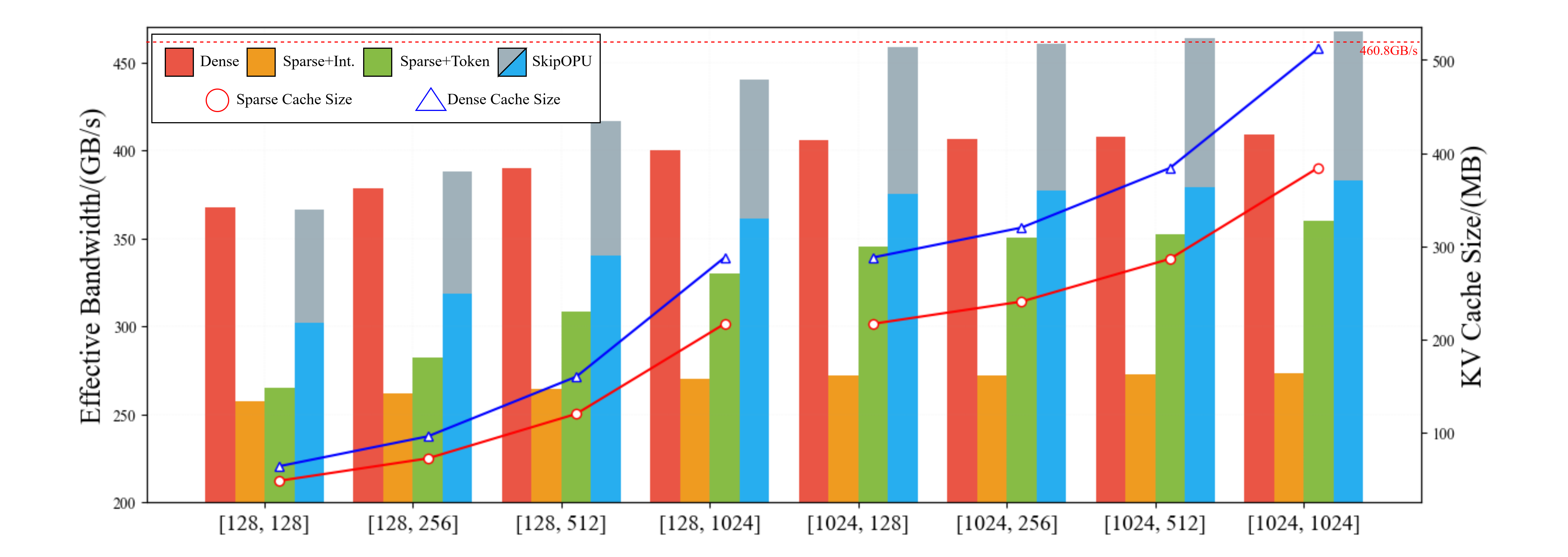}
    \caption{Effective memory bandwidth for KV cache with different mapping and scheduling options.}
    \label{fig:bandwidth}\vspace{-1.5em}
\end{figure}

\subsection{End-to-End performance comparison}
We further evaluate \textbf{SkipOPU} against state-of-the-art GPU and FPGA-based accelerators, with detailed results provided in Table \ref{tab:perform_comp}. 
This comparison accounts for diverse system characteristics, including hardware platforms, peak off-chip bandwidth, operating frequencies, and decoding configurations. 
\revC{
For a fair comparison, two different model scales are chosen and the decoding configuration within each benchmark is aligned as closely as possible with every baseline that specified its decoding configuration.
All state-of-the-art designs are evaluated under unified metrics—Normalized Throughput ($T_{norm}$) and Bandwidth (BW) Efficiency—to further mitigate variations caused by hardware platforms and static quantization methods. 
Specifically, $T_{norm}$ is defined as $T_{norm} = \frac{T\times M_{param}}{M_{calib}}$, where $T$ denotes the reported token throughput, $M_{param}$ represents the parameter size of the evaluated model and $M_{calib}$ represents the baseline parameter footprint of a 7B model with dense 4-bit quantization.
Concurrently, BW Efficiency refers to the ratio between the observed throughput $T$ and the theoretical decoding throughput can be achieved via the available memory bandwidth. 
Under the Llama2 workload, \textbf{SkipOPU} achieves $140.8$ Token/s and maintains $86.7\%$ and $89.1\%$ BW Efficiency for 7B and 13B models, respectively, yielding a $1.31\times - 2.55\times$ improvement over competing accelerators and GPUs. For the gpt2-345m task, our results are obtained by applying the same dynamic layer skipping method under $25\%$ sparsity ratio to the baseline model. The derived BW efficiency experiences a sharp drop to $48.8\%$ due to the inherent computation and control overheads that stall execution and data consumption, exert backpressure on the memory interface to force idleness and under-saturation. However, \textbf{SkipOPU} still outperforms alternative HBM-based designs like LoopLynx\cite{LoopLynx}($44.5\%$) and DFX \cite{DFX}($34\%$).
}

This global performance advantage is driven by superior HBM interface utilization and our dynamic routing mechanism, which elides redundant computations and memory accesses for unimportant tokens and layers.
Notably, while prior FPGA accelerators often report peak throughput under restricted decoding sequences, \textbf{SkipOPU} maintains its competitive performance even at a 1,024-token decode length. 
This scalability is a direct result of our KV-reuse mechanism and the KV-invariance buffer, which simultaneously reduces the HBM memory footprint and provides high aggregate bandwidth for sparse cache accesses.

\begin{table}[htp]
\centering
\small
\caption{End-to-end performance comparison between \textbf{SkipOPU} and other designs on different platforms}
\begin{threeparttable}
\begin{tabular}{ccccccccccccccccc}
\toprule
Design & \multicolumn{2}{c}{LoopLynx} & \multicolumn{2}{c}{MCoreOPU} & \multicolumn{2}{c}{Chen\cite{10.1145/3656177}} & \multicolumn{2}{c}{DFX} & \multicolumn{2}{c}{SkipOPU} & \multicolumn{2}{c}{vLLM} & \multicolumn{2}{c}{FlightLLM} & \multicolumn{2}{c}{SkipOPU} \\
\midrule
Device & \multicolumn{2}{c}{U50} & \multicolumn{2}{c}{U200} & \multicolumn{2}{c}{U280} & \multicolumn{2}{c}{U280} & \multicolumn{2}{c}{U280} & \multicolumn{2}{c}{A100} & \multicolumn{2}{c}{U280} & \multicolumn{2}{c}{U280} \\
Bandwidth(GB/s) & 201 & 402 & \multicolumn{2}{c}{76.8} & \multicolumn{2}{c}{460} & \multicolumn{2}{c}{460} &  \multicolumn{2}{c}{460} & \multicolumn{2}{c}{1555} & \multicolumn{2}{c}{460} & \multicolumn{2}{c}{460} \\
\midrule
Frequency(MHz) & 285 & 285 & \multicolumn{2}{c}{300} & \multicolumn{2}{c}{245} & \multicolumn{2}{c}{200} & \multicolumn{2}{c}{225(450)}  &  \multicolumn{2}{c}{1410} & \multicolumn{2}{c}{225} & \multicolumn{2}{c}{225(450)} \\
Tasks & \multicolumn{2}{c}{gpt2-245m} & \multicolumn{2}{c}{gpt2-345m} & \multicolumn{2}{c}{gpt2-345m}& \multicolumn{2}{c}{gpt2-345m} & \multicolumn{2}{c}{gpt2-345m} & \multicolumn{2}{c}{llama2-7b} & \multicolumn{2}{c}{llama2-7b} & {\footnotesize \makecell{llama2\\-7b}} & {\footnotesize \makecell{llama2\\-13b}} \\
Opt. for W & \multicolumn{2}{c}{W8} &\multicolumn{2}{c}{W8} & \multicolumn{2}{c}{W8} & \multicolumn{2}{c}{HF16} & \multicolumn{2}{c}{W4} & \multicolumn{2}{c}{HF16} & \multicolumn{2}{c}{SparseW8} & \multicolumn{2}{c}{W4} \\
$[\text{prefill, decode}]$ & \multicolumn{2}{c}{N/A} & \multicolumn{2}{c}{[128, 256]} & \multicolumn{2}{c}{N/A} & \multicolumn{2}{c}{[128, 256]} & \multicolumn{2}{c}{[128, 256]} & \multicolumn{2}{c}{[128, 1024]} & \multicolumn{2}{c}{N/A}&  \multicolumn{2}{c}{[128, 1024]} \\
Token/s & 260 & 392 & \multicolumn{2}{c}{45.0} & \multicolumn{2}{c}{204} & \multicolumn{2}{c}{124.1} & \multicolumn{2}{c}{1427.4} & \multicolumn{2}{c}{45.3} & \multicolumn{2}{c}{55} & 140.8 & 75.6 \\
Norm Throughput & 25 & 37.7 & \multicolumn{2}{c}{4.3} & \multicolumn{2}{c}{19.6} & \multicolumn{2}{c}{23.8} & \multicolumn{2}{c}{79.3} & \multicolumn{2}{c}{181.2} & \multicolumn{2}{c}{55} & 140.8 & 144.5\\ 
BW efficiency & 59\% & 44.5\% & \multicolumn{2}{c}{70\%} & \multicolumn{2}{c}{23\%} & \multicolumn{2}{c}{34\%} & \multicolumn{2}{c}{48.8\%} & \multicolumn{2}{c}{31.5\%} &  \multicolumn{2}{c}{66\%} & 86.7\% & 89.1\% \\
\bottomrule
\end{tabular}
\end{threeparttable}
\label{tab:perform_comp}
\end{table}

\begin{figure}
    \centering
    \includegraphics[width=0.9\textwidth]{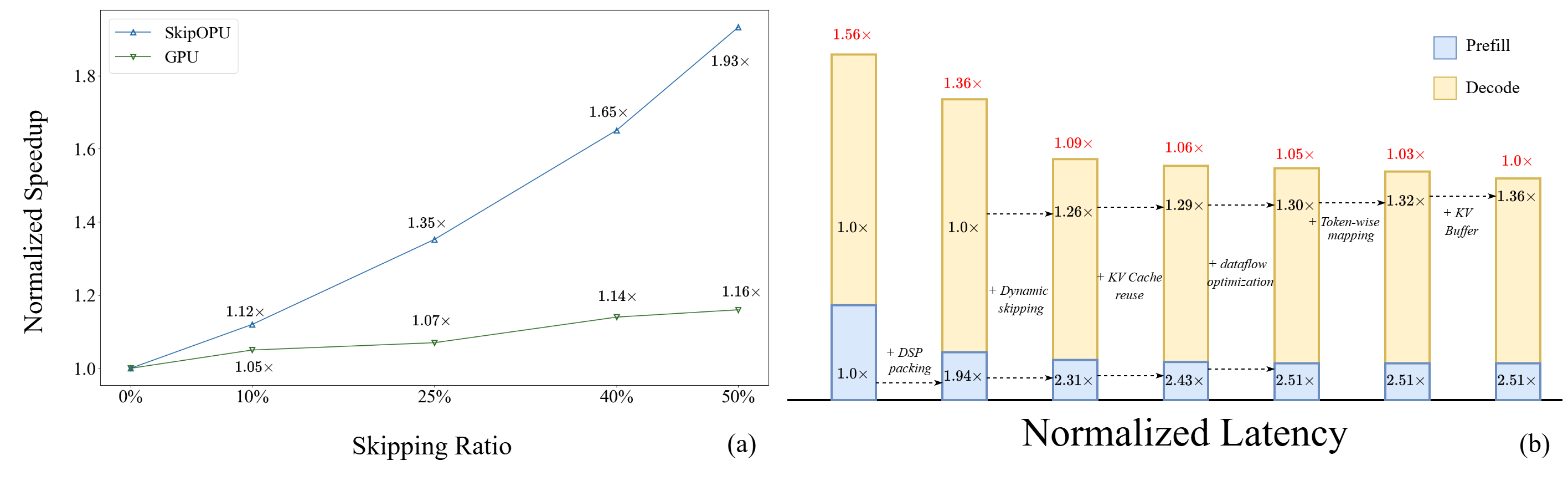}
    \caption{Comprehensive performance evaluation of \textbf{SkipOPU}.(a) Sensitivity analysis to the skipping ratio on SkipOPU and GPU. (b) End-to-end speedup breakdown }
    \label{fig:analysis}\vspace{-1.5em}
\end{figure}

\revC{
\subsection{Performance Evaluation}
In this section, we conduct an end-to-end performance evaluation comprising two dimensions: an algorithmic sensitivity analysis across varying skipping intensities, and a speedup breakdown on the Llama2-7b with $25\%$ skipping ratio under $[512, 1024]$ decoding configuration. \Cref{fig:analysis} illustrates these system-level performance characteristics.

\textit{Sensitivity Analysis.} \Cref{fig:analysis}.(a) illustrates the normalized system speedup as a function of the dynamic skipping ratio ($\alpha$). 
As shown, the performance of \textbf{SkipOPU} scales almost linearly with $1/(1-\alpha)$, demonstrating that our architecture successfully translates the reduction in operational intensity into proportional performance gains. 
Notably, at relatively low skipping ratios ($\alpha \le 25\%$), the derived speedup actually exceeds the theoretical algorithmic operation-reduction curve. 
This phenomenon stems directly from the micro-architectural synergy between the token-wise HBM mapping scheme and the on-chip $KV$ invariance buffer. 
Together, these structures optimize spatial access locality and introduce hierarchical data reuse, yielding an aggregate effective bandwidth that surpasses the baseline dense HBM performance, thereby further accelerating the MHA subsystem throughput beyond $1/(1-\alpha)\times$. 
Conversely, under high-sparsity scenarios ($\alpha \ge 40\%$), the drastically increased volume of inter-layer skipped data forces a significantly higher rate of read and write-back transactions to the on-chip $KV$ invariance buffer. 
Once this transaction volume exceeds the physical access bandwidth capacity of the instantiated URAMs, the fundamental ``zero-overhead'' prerequisite of our memory hierarchy is violated. 
The resulting structural hazards on the buffer ports induce execution pipeline stalls, degrading the aggregate memory bandwidth and causing the overall speedup to plateau below the theoretical algorithmic gain. 

\textit{Speedup Breakdown.} \Cref{fig:analysis}.(b) depicts the a normalized latency breakdown across cumulative optimization stages, tracking the prefill and decode execution phases independently, where the annotation within each bin represents the throughput improvement over the baseline. 
The baseline system exhibits a total normalized latency of $1.56\times$. 
Integrating our hardware \emph{DSP packing} mechanism heavily accelerates the compute-bound prefill stage, compressing its latency by $1.94\times$ via parallelized arithmetic execution. 
The \emph{dynamic skipping} method proportionally reduces both the computational intensity during the prefill stage and the memory transactions required during the decode stage, yielding corresponding throughput improvements of $1.19\times$ ($2.31/1.94$) and $1.26\times$, respectively. The \emph{KV Cache reuse} strategy further deduce the computation and memory transactions associated with those unimportant tokens, and the total latency is reduced to $1.06\times$. The \emph{dataflow optimization} is applied to eliminate the execution stalls between the linear and nonlinear operations, which improves both the prefill and decode throughput by $8\%$ and $1\%$, respectively. To mitigate the severe cross-layer row-buffer thrashing introduced by \emph{KV Cache reuse} strategy, \emph{token-wise mapping} is implemented for active HBM KV entries to reduce memory fragmentation and reclaims spatial access locality, achieving a $1.03\times$ total latency. Finally, instantiating the full \emph{$KV$ invariance buffer} absorbs the remaining cross-layer channel interference, bringing the system to its optimal $1.0\times$ normalized end-to-end execution latency, with a total $2.51\times$ and $1.36\times$ throughput improvement over the prefill and decode stage of the baseline, respectively.
}

\section{Related Work}
\textbf{FPGA-based LLM Accelerators.} 
Existing FPGA accelerators for transformers generally assume a fixed execution graph \cite{FET-OPU, DFX, 10.1145/3564606, PP-Transformer, 10818746, METAL, MEADOW}. 
These designs target either dense models or static parameter sparsity, and can only improve hardware efficiency for predetermined computation patterns.
Some recent works begin to incorporate runtime adaptivity, such as leveraging token relevance within the attention mechanism \cite{SPATTEN}. However, they focus on sparsity within a single structural dimension and do not capture the heterogeneous importance of tokens across both layers and sequences, leaving substantial redundancy unaddressed.
Our design explicitly supports token-wise and layer-wise dynamic execution, enabling the hardware to adapt its computation and memory behavior to the model’s runtime decisions and thereby providing high-performance inference for LLMs on FPGA. 

\revA{
\textbf{DSP Packing.}
To mitigate memory and logic constraints in deep learning accelerators, industry-standard FPGA implementations pioneered low-precision arithmetic packing within primitive DSP blocks to maximize throughput per unit area~\cite{xilinx_whitepaper_2017, xilinx_whitepaper_2020}.
However, these vendor techniques target specific arithmetic requirements and lack the versatility to adapt to varied bit-widths.
Academic frameworks addressed this rigidity by generalizing multiplication packing through overpacking methodologies~\cite{Overpack}, successfully improving DSP block utilization across both fixed-point computation~\cite{luo2023deepburning, LI2026103627} and floating-point mantissa calculations~\cite{FlightVGM}.
Despite maximizing multiplier utilization, these methods treat multiplication as a monolithic block and use only a minimal fraction of the internal 48-bit post-adder for error correction, forcing expensive post-processing logic onto the external LUTs ~\cite{DWA}.
\textbf{SkipOPU} circumvents this structural limitation by decomposing parallel multiplications into successive partial addition stages, selectively truncating operand bit-widths to minimize guard-bit corruption overhead while the residual information lost is preserved and routed via the DSP's native $C$-port to fully utilize the internal adder fabric.
}

\textbf{Nonlinear Processing Engine.} Nonlinear operations have been widely recognized as a major bottleneck in LLM inference due to their low arithmetic intensity and strong data dependencies. To alleviate this issue, VPE \cite{VPE} introduces a deeply pipelined dataflow that reduces pipeline stalls caused by dependency chains. However, the row-oriented processing pattern in VPE is mismatched with the tile-stream outputs of typical PE arrays, which introduces unavoidable buffering latency when interfacing linear and nonlinear stages.
While DESA \cite{DESA} and METAL \cite{METAL} both utilize decoupled strategies to overlap nonlinearities, DESA relies on aggressive spatial parallelism that leads to substantial hardware overhead and underutilization. In contrast, our NPE not only maintains a resource-efficient throughput-matching design but significantly expands the functional scope beyond the BERT-focused operations of prior works. By adopting a unified tile-wise dataflow, we seamlessly integrate modern LLM-specific components like RoPE and SwiGLU into the execution pipeline. 

\textbf{KV Cache Management.} vLLM \cite{vLLM} alleviates memory fragmentation by organizing KV tensors in paged storage, whereas ChatOPU \cite{ChatOPU} improves effective off-chip bandwidth by enforcing regular AXI access patterns through coalesced spill and reload operations.
Both approaches implicitly rely on relatively predictable KV access behavior. In contrast, dynamic layer skipping introduces cross-layer KV reuse, which results in irregular and contention-prone memory transactions. By exploiting the invariance of reused KV entries and maintaining them on chip, our design converts fragmented HBM accesses into localized reuse while preserving burst transfers for newly generated KV data.

\section{Conclusion}
This paper proposes \textbf{SkipOPU}, an FPGA-based overlay processor for LLMs with dynamically allocated computation across both tokens and layers. 
At the algorithmic level, we employ KV reuse strategy for KV entries of those skipped tokens to completely bypass the MHA computation under the SkipGPT framework and further apply the fused dataflow of the router and the self-attention mechanism to facilitate the latency hiding of nonlinearities. 
At the hardware level, 
we develop a mixed precision processing engine with fine-grained packing to fully exploit the inherent DSP arithmetic capability, while the accumulation overhead is minimized through one-time FP-BFP conversion and lightweight fixed-point summation trees. 
In addition, cross-layer KV invariance is leveraged to proactively buffer reusable KV entries on-chip, which regularizes otherwise randomized HBM accesses and augments the aggregate memory throughput with on-chip bandwidth supply.
Experiments demonstrate that our \textbf{SkipOPU} implemented on Xilinx U280 platform achieves $1.31\times$–$2.55\times$ higher bandwidth efficiency than the state-of-the-art GPU application and other FPGA-based accelerators while reducing KV cache storage by up to 25.4\% across varying input and output sequence lengths.

\bibliographystyle{ACM-Reference-Format}
\bibliography{reference}

\end{document}